\documentclass[11pt,a4paper]{article}
\pdfoutput=1
\usepackage{jheppub}
\usepackage{comment}
\usepackage{gensymb}
\usepackage{longtable}
\usepackage{amssymb}
\usepackage{lipsum}
\DeclareSymbolFont{matha}{OML}{txmi}{m}{it}
\DeclareMathSymbol{\varv}{\mathord}{matha}{118}
\usepackage{amsmath}
\usepackage{amssymb}
\usepackage{graphicx}
\usepackage{subfigure}
\usepackage{pstricks}
\usepackage{bm}
\usepackage{lscape}
\usepackage{pbox}
\usepackage{placeins}
\usepackage{booktabs}
\usepackage[T1]{fontenc}
\usepackage{footnote}
\usepackage{pdfpages}
\usepackage{hhline}
\usepackage{multirow}
\usepackage{multicol}
\usepackage{enumitem}
\usepackage[toc,page]{appendix}
\allowdisplaybreaks
\usepackage{comment}
\usepackage{float}                    
\usepackage{slashed}
\usepackage{mathrsfs}

\usepackage{color, colortbl}
\definecolor{LightCyan}{rgb}{0.88,1,1}

\usepackage{soul}
\setstcolor{red}
\usepackage{textcomp}
\usepackage{gensymb}
\usepackage{xcolor,pifont}

\newcommand*\colourcheck[1]{%
  \expandafter\newcommand\csname #1check\endcsname{\textcolor{#1}{\ding{52}}}%
}
\colourcheck{blue}
\colourcheck{green}
\colourcheck{red}
\colourcheck{olivegreen}

\usepackage{empheq}
\usepackage[numbers]{natbib}
\usepackage{notoccite}

\usepackage{rotating}
\usepackage{tabularx}
\usepackage[labelfont=bf]{caption} 

\FloatBarrier

\renewcommand{\thetable}{\arabic{table}}

\newcommand{\beq}{\begin {equation}}  
\newcommand{\eeq}{\end   {equation}} 
\newcommand{\bea}{\begin {eqnarray}} 
\newcommand{\eea}{\end   {eqnarray}}  
\newcommand{\baa}{\begin {array}   } 
\newcommand{\eaa}{\end   {array}   }     
\newcommand{\bit}{\begin {itemize} }
\newcommand{\eit}{\end   {itemize} }
\newcommand{\be }{\begin {equation}} 
\newcommand{\ee }{\end   {equation}}

\definecolor{MyDarkBlue}{rgb}{0.1, 0.1, 0.8} 
\definecolor{SBlue}{rgb}{0.2, 0.4, 0.7} 
\definecolor{MyLightBlue}{rgb}{0.22,0.51,0.9}
\definecolor{MyGreen}{rgb}{0.0, 0.5, 0.3}
\definecolor{BrickRed}{rgb}{0.8, 0.25, 0.33}
\RequirePackage{hyperref}
\hypersetup{colorlinks, citecolor=BrickRed,linkcolor=MyDarkBlue, urlcolor=MyLightBlue}

\begin{document}
\preprint{MS-TP-24-24}
\title{\textbf{Chiral dark matter and radiative neutrino masses from gauged \boldmath{$U(1)$} symmetry}}
 \author[a]{K.S. Babu,}
 \author[b]{Shreyashi Chakdar,}
 \author[c]{and Vishnu P.K.}

\affiliation[a]{Department of Physics, Oklahoma State University, Stillwater, OK 74078, USA}
\affiliation[b]{Department of Physics, College of the Holy Cross, Worcester, MA 01609, USA}
\affiliation[c]{Institut f{\"u}r Theoretische Physik, Universit{\"a}t M{\"u}nster, Wilhelm-Klemm-Stra\ss{}e 9, 48149 M{\"u}nster, Germany}
 
\emailAdd{babu@okstate.edu, schakdar@holycross.edu, vishnu.pk@uni-muenster.de}
\abstract{We propose a class of dark matter models based on a chiral $U(1)$ gauge symmetry acting on a dark sector. The chiral $U(1)$ protects the masses of the dark sector fermions, and also guarantees the stability of the dark matter particle by virtue of an unbroken discrete $\mathcal{Z}_N$ gauge symmetry. We identify 38 such $U(1)$ models which are descendants of a chiral $SU(3) \times SU(2)$ gauge symmetry, consisting of a minimal set of fermions with simple $U(1)$ charge assignments. We show how these models can also be utilized to generate small Majorana neutrino masses radiatively via the scotogenic mechanism with the dark sector particles circulating inside loop diagrams. We further explore the phenomenology of the simplest model in this class, which admits a Majorana fermion, Dirac fermion or a scalar field to be the dark matter candidate, and show the consistency of various scenarios with constraints from relic density and direct detection experiments.}

\maketitle
\setcounter{footnote}{0}

\section{Introduction}\label{SEC-01}

An understanding of the nature of dark matter (DM) in the universe as well as the origin of small neutrino masses are two of the glaring limitations of the highly successful Standard Model (SM) of elementary particles.  A variety of models that go beyond the Standard Model have been explored to address these issues. The existence of a dark sector has been postulated to address the nature of dark matter, populated by new particles and possibly new gauge symmetries~\cite{Cooley:2022ufh}.  Such a dark sector could be secluded from the SM sector~\cite{Pospelov:2007mp}, enabling the identification of the lightest stable particle from that sector to be a dark matter candidate. 
A simple example of such a setup is provided by a dark $U(1)_D$ gauge symmetry consisting of a dark photon, dark fermions and a dark scalar.  Controlled cross-couplings, such as Higgs mixing between the two sectors, or photon-dark photon kinetic mixing, can result in satisfying the correct relic abundance of DM. Typically such models assume that the dark sector fermion(s) $\chi$, transform vectorially under the $U(1)_D$ gauge symmetry, in which case a bare mass term $m_\chi \overline{\chi} \chi$ would be permitted in the Lagrangian.  Such a bare mass is not protected by any symmetries and can take large values, as large as the Planck scale.  Consistency with DM phenomenology would require, however, that $m_\chi$ should be not above a few tens of TeV. 
It would seem, therefore, to be more natural if the fermions from the dark sector acquire their masses only after spontaneous $U(1)_D$ breaking when a scalar field $\eta$ acquires a nonzero vacuum expectation value, through a Yukawa interaction term of the type $y_\chi \overline{\chi} \chi \eta$ in the Lagrangian.  The mass of $\chi$, $m_\chi = y_\chi \langle \eta \rangle$, is protected in this case down to the gauge symmetry breaking scale, just as the top quark mass is protected in the SM down to the electroweak symmetry breaking scale by the SM gauge symmetry. 

In order to prevent the bare mass term for $\chi$ and to generate its mass spontaneously, the $U(1)_D$ gauge theory should be a {\it chiral theory}, just as the electroweak theory.  The fermionic dark matter arising from such theories are termed {\it chiral dark matter}~\cite{Abe:2019zhx,Berezhiani:1995yi,Berezhiani:2000gw,
Foot:1991bp,Foot:2004pa,Berryman:2016rot,BhupalDev:2016gna,Berryman:2017twh,Dror:2020jzy,Babu:2023zni,Abe:2023yte}. In spite of its attractive features, very few models are available in the literature realizing such chiral dark matter candidates. The requirement of anomaly cancellation sets strong constraints on any chiral $U(1)$ gauge theory.  For example, there should be a minimum of five Weyl fermions in any such model in order to cancel the gauge anomalies~\cite{Babu:2003is,Batra:2005rh,Sayre:2005yh,Costa:2019zzy,Costa:2020dph}, provided that the $U(1)$ charges are rational numbers. One of the main objectives of this paper is to provide a class of chiral $U(1)$ models that can be adopted as the dark sector gauge theory.  This class of models has its origin in a chiral $SU(3) \times SU(2)$ theory, which is one of the simplest chiral non-Abelian gauge theories, with one family of fermions resembling a single family of the SM. At some high energy scale $SU(3) \times SU(2)$ gauge symmetry breaks spontaneously down to $U(1)_D$, which in the process removes some of the chiral fermions in the original theory by giving them spontaneously induced masses. Our results for the particle content of the surviving $U(1)_D$ models are summarized in Table \ref{table_solutions}, where we have identified 38 simple models. These models are characterized by having at most 8 Weyl fermions with their $U(1)_D$ charges being integers, which are restricted so that none of the fermion charge ratios exceed 7 in magnitude. In all models listed, but for the first three (which have fewer Weyl fermions), a single scalar field is sufficient to generate masses for the dark matter as well as for its fermionic partners.  In this sense, these are minimal chiral models. We also carry out the phenomenology of the simplest such DM models from this list, namely, Models 4 and 5 of Table \ref{table_solutions}. An interesting byproduct of this model building exercise is that an unbroken $\mathcal{Z}_N$ symmetry survives even after the spontaneous breaking of $U(1)_D$, which guarantees the stability of the dark matter. In the explicit models that we study, viz., Models 4 and 5, this symmetry is $\mathcal{Z}_2$ and $\mathcal{Z}_4$, respectively. 

The second glaring limitation of the Standard Model, viz., an understanding of the origin of small neutrino masses, may be addressed in the same setup that explains the nature of dark matter.  This is achieved through the so-called scotogenic mechanism~\cite{Ma:2006km}, wherein small Majorana neutrino masses are induced through loop diagrams in which dark sector particles propagate~\cite{Tao:1996vb}. Neutrino masses are small, primarily because they arise via loops. We implement this mechanism in two of our models given in Table \ref{table_solutions}.  (For alternative methods of implementing scotogenic mechanism, see Ref.~\cite{Krauss:2002px,Aoki:2008av,Ma:2008cu,Restrepo:2013aga,Gustafsson:2012vj,Fiaschi:2018rky,Esch:2018ccs,Ma:2013yga,Ho:2016aye,Wong:2020obo,deBoer:2023phz,Herms:2023cyy,Jana:2024iig}.) This requires introduction of two additional scalar fields into these theories. The scalar particle content that we adopt for Models 4 and 5 are listed in Table \ref{models}.  Since this mechanism requires the existence of some fields connecting the neutrinos with the dark sector (e.g., the $\Phi$ field in Table \ref{models}), the scotogeneic models do not fall into purely dark sector models.  However, given the smallness of neutrino masses, it turns out that such interlocking terms do not play any important role in the dark matter phenomenology that we study, and we shall continue to refer to these models as dark sector models.

Our main results can be summarized as follows. We have constructed a class of chiral $U(1)$ gauge models which can be employed as the dark sector containing a stable dark matter candidate.  The 38 models with reasonable charge ratios that originate from a non-Abelian $SU(3) \times SU(2)$ chiral theory are listed in Table \ref{table_solutions}.  A single scalar is sufficient to generate masses for all dark sector particles in these models, except for the first three, which have fewer Weyl fermions.  We have also implemented the scotogenic mechanism for neutrino mass generation in two of the simplest models of Table \ref{table_solutions}.  Furthermore, we have studied the theoretical and experimental constraints on these models and analyzed their dark matter phenomenology. It turns out that when the scotogenic mechanism is implemented, there are more than one option for the dark matter candidate, which may be a Majorana fermion that directly takes part in the neutrino mass generation, a Dirac fermion which is an anomaly-cancelling partner of the Majorana fermion, or a scalar field which is a mixture of an $SU(2)_L$ doublet and a singlet. We have explored the DM phenomenology for each scenario and have shown their consistency with relic abundance as well as direct detection limits.

We believe that the models presented here are some of the simplest  chiral fermion dark matter models that have several appealing features.  In particular, these models shed light on two of the puzzling features of the prototypical scotogenic models, namely, the origin of a $\mathcal{Z}_2$ symmetry put in by hand in order to stabilized the DM, and the mass scale of the dark sector fermion. Our models have a surviving $\mathcal{Z}_N$ gauge symmetry originating from $U(1)_D$ that ensures DM stability, and the gauge symmetry protects the mass of dark sector fermions, including that of the dark matter candidate.

The rest of the paper is organized as follows. In Sec. \ref{SEC-02} we identify simple chiral dark matter models originating from a $SU(3) \times SU(2)$ theory.  In Sec. \ref{SEC-03} we develop the phenomenology of the simplest of these models.  In Sec. \ref{SEC-04} we analyze the dark matter phenomenology of the model and show the consistency of various scenarios for dark matter candidates (Majorana or Dirac fermion, as well as doublet-singlet DM candidates). Finally, in Sec. \ref{SEC-05} we conclude.

\section{Chiral \boldmath{$U(1)_D$} models originating from \boldmath{$SU(3) \times SU(2)$}}\label{SEC-02}
In this section, we describe the method we followed to derive a class of chiral $U(1)_D$ gauge models to be identified as the dark sector of Standard Model extensions. Any chiral gauge theory is constrained by the requirement of anomaly cancellation.  We use these conditions to restrict the particle content of the model, as well as the $U(1)_D$ charges of various fermions within a specific framework. We focus on  SM $\times \,U(1)_D$ extensions wherein all SM particles are neutral under $U(1)_D$. 
Denoting the $U(1)_D$ charges of a generic dark sector fermion as $q_i$ with $i=1-M$ where $M$ is the number of left-handed Weyl fermions in the theory, the anomaly cancellation conditions are
\begin{align}
&\sum_{i=1}^{M} q_i = 0, \label{eq:2.1} \\  
&\sum_{i=1}^{M} q_i^3 = 0 \label{eq:2.2},
\end{align}
where the condition (\ref{eq:2.1}) derives from the mixed gauge-gravitational anomaly while the condition (\ref{eq:2.2}) derives from the cubic gauge anomaly.  In general, these equations have infinite number of solutions, if the number of Weyl fermion $M$ is left unspecified.  Our goal here is to identify solutions which are chiral with the smallest number of fermion fields. If a solution has $q_k = -q_l$, then the fermions with charges $q_k$ and $q_l$ are vectorial, since a bare mass connecting them is permissible.  In identifying chiral gauge theories we would remove such vector-like pairs from the theory. All charges should be real numbers for consistency; we further require them to be also rational numbers, in which case they can be taken to be integers by a rescaling.

We seek solutions to the anomaly cancellation conditions of Eqs. (\ref{eq:2.1})-(\ref{eq:2.2}) that arise from the chiral fermions of a simple non-Abelian gauge theory based on $SU(3) \times SU(2)$.  The chiral fermion spectrum of this theory is identical to the one-family Standard Model fermion spectrum with the hypercharge analog eliminated.  This theory is one of the simplest chiral gauge theories.  A chiral $U(1)_D$ that originates from this theory upon spontaneous symmetry breaking would automatically be anomaly free.  The fermion spectrum of the $SU(3) \times SU(2)$ theory can be written, in analogy with the SM fermions, as
\begin{eqnarray}
&~&Q'(3,2) = \left( \begin{matrix}
u_1' & u_2' & u_3' \cr d_1' & d_2' & d_3' 
\end{matrix} \right),~~~~~L'(1,2)=\left( \begin{matrix} \nu' \cr \ell' 
\end{matrix} \right) \nonumber \\
&~&\bar{U}'(3^*,1)=(\begin{matrix} \bar{u}'_1 & \bar{u}_2' & \bar{u}_3' \end{matrix}),~~~~~~\bar{D}'(3^*,1)=(\begin{matrix} \bar{d}_1' & \bar{d}_2' & \bar{d}_3' \end{matrix})~.\label{spectrum}
\end{eqnarray}
Here all fermion fields are taken to be left-handed.
This is an anomaly-free chiral theory  since no bare mass terms are allowed for any of these fermions.  The $L'$ field is needed here in order to cancel the global Witten anomaly which requires the number of $SU(2)$ doublets to be even~\cite{Witten:1982fp}. Now, this gauge symmetry has three diagonal generators, parameterized by the Gell-Mann matrices $\lambda_3$ and $\lambda_8$ and the Pauli matrix $\sigma_3$ corresponding to $SU(3)$ and $SU(2)$, respectively.  Any linear combination of $U(1)$ charges proportional to these diagonal generators will lead to an anomaly-free $U(1)_D$ symmetry, which in most instances will have chiral fermions.  We adopt this approach to explore chiral $U(1)_D$ models. The components of $L'$ in Eq. (\ref{spectrum}) would have equal and opposite charges under such a $U(1)_D$ and therefore will not contribute to the anomalies.  We are then left with 12 Weyl fermions in the theory.  with their charges  $\{q_i\}$ given as
\begin{align}
     & \left(\{a+b+c\}, \{a+b-c\}, \{a-b+c\}, \{a-b-c\},\{-2a+c\}, \{-2a-c\}, \right. \nonumber \\ 
     & \,\, \left.  \{-a-b\},\{-a-b\}, \{-a+b\},\{-a+b\},\{2a\},  \{2a\}\right)
     \label{charge}
\end{align}
where $a, b$ and $c$ are arbitrary real numbers corresponding to the $\lambda_8$, $\lambda_3$ and $\sigma_3$ generators, respectively. We then check for chiral solutions to the anomaly cancellation conditions of Eqs. (\ref{eq:2.1})-(\ref{eq:2.2}) with the minimal number of fermions, $M < 12$, for non-zero values of $a, b,c$. The number of fermion fields being $M < 12$ is realized by removing those fields which have vectorial charges.

In Table.~\ref{table_solutions}, we have listed a class of simple anomaly-free chiral models obtained within this setup. To obtain this Table, we wrote a computer program and randomly scanned $a,b,c$ over integers in a range $\left[-15,15\right]$ for $10^6$ times. From the obtained pool of solutions we chose the 38 chiral models satisfying the added restrictions that the number of chiral fermions be $M \leq 8$ and the magnitude of the maximal charge ratio obey $|q_i/q_j|\leq 7$ for all $(i,j)$.
Additionally, for each of these chiral model, we have investigated the minimal set of scalar fields that can give masses for all the chiral fermions and have presented the corresponding charge assignments of the scalar fields in the last column of the Table.~\ref{table_solutions}. 
It is to be noted that out of the the 38 obtained solutions with the restrictions explained above, two solutions have six  chiral fermions, one solution has seven chiral fermions and the rest of the solutions have eight fermions. However, the three solutions with six and seven fermions require at least two Higgs fields to generate masses for all the new fermions, whereas the remaining 35 solutions with eight chiral fermions require only one Higgs field. As a result, we choose two models (Models 4 and 5 in the Table ~\ref{table_solutions}) with eight chiral fermions and one Higgs field that can generate fermion masses as our benchmark models. We perform a detailed dark matter phenomenology for one of these benchmark models, Model 4, in Secs. \ref{SEC-03}-\ref{SEC-04}.  Model 5 has a very similar phenomenology, which we also outline in these sections.

It is instructive to see how these models emerge from $SU(3) \times SU(2)$.  Consider Model 4 as an example. At the $SU(3) \times SU(2)$ level, the unbroken $U(1)_D$ charges of the fermions of Eq. (\ref{spectrum}) for Model 4 are given by
\begin{eqnarray}
&~&[Q'(3,2)] = \left( \begin{matrix}
-5 & -3 & ~2 \cr -1~ & ~1 & 6 
\end{matrix} \right),~~~~~[L'(1,2)]=\left( \begin{matrix} -2 \cr ~2 
\end{matrix} \right) \nonumber \\
&~&[\bar{U}'(3^*,1)]=(\begin{matrix} 3 & 1 & -4 \end{matrix}),~~~~~~[\bar{D}'(3^*,1)]=(\begin{matrix} 3 & 1 & -4 \end{matrix})~.
\end{eqnarray}
It is clear that $u_2'$ can pair up with $\bar{u}_1'$, $d_1'$ can pair with $\bar{d}_2'$ and $u_3'$ with $\nu'$, leaving behind 8 chiral fermions, with charges given as in Model 4 of Table~\ref{table_solutions}.  

This example also illustrates possible ways of breaking $SU(3) \times SU(2)$ gauge symmetry down to the desired $U(1)_D$ symmetry.  The pairing of $u_2'$ with $\bar{u}_1'$ as well as $d_1'$ with $\bar{d}_2'$ can be achieved by a Higgs field transforming as $(8,2)$, denoted as $\Phi_{i\alpha}^\beta$, satisfying the condition $\sum_{\alpha=1}^3 \Phi_{i\alpha}^\alpha = 0$. Here $(\alpha,\,\beta)$ are $SU(3)$ indices while $i$ is $SU(2)$ index. The Yukawa coupling $Q'_{i \alpha} \bar{U}'^\beta \Phi_{j\beta}^\alpha \epsilon^{ij}$
and $Q'_{i \alpha} \bar{D}'^\beta \Phi_{j\beta}^\alpha \epsilon^{ij}$ will generate these Dirac masses once the components $\Phi_{12}^1$ and $\Phi_{21}^2$ acquire nonzero vacuum expectation values.  Similarly, a mass term connecting $u_3'$ and $\nu'$ will arise from the Yukawa coupling involving a Higgs field $\Delta(3^*, 3)$, 
 $Q'_{i\alpha} L'_j \Delta^{i \alpha}_k \epsilon^{jk}$,  where $\sum_{i=1}^2\Delta_i^{i \alpha} = 0$,
when $\Delta_2^{13}$ acquires a nonzero VEV.  It can be checked that $SU(3) \times SU(2)$ breaks down to $U(1)_D$ with these nonzero VEVs. 

While the models of Table~\ref{table_solutions} were obtained by numerically scanning for integers $(a,b,c)$ in a specific range, it should be mentioned that this is a finite set, for the number of fermions being 8 or less. Among the 12 fermion charges given in Eq. (\ref{charge}), one can set any two charges to be equal and opposite to give mass and thus remove a pair of fermions. When the $U(1)_D$ theory has 8 fermions, two conditions have to be satisfied: $q_i + q_j = 0$ and $q_k+q_l=0$ for differing $(i,j,k,l)$.  These two conditions will determine two of the parameters in the set $(a,b,c)$, and thus the model is defined, since the overall charge is arbitrary.  Of the 12 fermionic charges of  Eq. (\ref{charge}), only 9 are independent owing to the degeneracy of $\overline{D}'$ and $\overline{U}'$ charges.  This leaves $9!/(7! \,2!)=36$ choices to pick the first pair of charges, and then $7!/(5! \,2!)=21$ choices to pick a pair from the remaining 7 charges.  The total number of possibilities is then $36 \times 21 = 756$.  The cases with 6 and 7 fermions require an additional condition, for which in general there is no solution.  Modes 1, 2, 3 are thus special solutions. Finally, if the number of fermions under the surviving $U(1)_D$ is 10 or more, there would be fewer conditions than parameters, which would lead to a family of solutions.

\FloatBarrier
\begin{table}[t!] 
\centering
\footnotesize
\resizebox{1.1\textwidth}{!}{
\begin{tabular}{|c|c|c|c|c|}
\hline
\pbox{10cm}{
\vspace{2pt}
Model  
\vspace{3pt}
}
 & 
\pbox{10cm}{
\vspace{2pt}
Number of \\
\centering Fermions
\vspace{3pt}
}
 & 
 \pbox{10cm}{
\vspace{2pt}
$(a,b,c)$  
\vspace{3pt}
}
 & 
 \pbox{10cm}{
\vspace{2pt}
Fermion States: \\
\centering Multiplicity$\times\{$charge$\}$ 
\vspace{3pt}
} & 
 \pbox{10cm}{
\vspace{2pt}
Scalar States: \\
\centering Multiplicity$\times\{$charge$\}$ 
\vspace{3pt}
} 
\\\hline
1 & 6 & $(-1,0,-3)$ & $3\times \{1\}+2\times \{-4\}+1\times \{5\}$ &  $1\times\{3\}+1\times\{6\}$
\\\hline
2 & 6 & $(-1,-4,-1)$ & $1\times \{1\}+1\times \{-2\}+1\times \{-3\}+2\times \{5\}+1\times \{-6\} $&  $1\times\{1\}+1\times\{2\}$
\\\hline
3 & 7 & $(-1,-2,-1)$ & $1\times \{-1\}+2\times \{-2\}+3\times \{3\}+1\times \{-4\}$ & $1\times\{1\}+1\times\{2\}$
\\\hline
\rowcolor{LightCyan}
4 & 8 & $(-2,-1,-2)$ & $2\times \{1\}+1\times \{2\}+1\times \{3\}+2\times \{-4\}+1\times \{-5\}+1\times \{6\}$ & $1\times\{2\}$ 
\\\hline
\rowcolor{LightCyan}
5 & 8 & $(-1,-2,-4)$ & $1\times \{-1\}+3\times \{-2\}+1\times \{3\}+1\times \{5\}+1\times \{6\}+1\times \{-7\}$ & $1\times\{4\}$
\\\hline
6 & 8 & $(-2,-5,-1)$ & $1\times \{2\}+1\times \{-3\}+1\times \{-4\}+1\times \{5\}+1\times \{-6\}+2\times \{7\}+1\times \{-8\}$ & $1\times\{1\}$
\\\hline
7 & 8 & $(-1,-4,-7)$ & $1\times \{-2\}+2\times \{-3\}+1\times \{-4\}+1\times \{5\}+1\times \{9\}+1\times \{10\}+1\times \{-12\}$ & $1\times\{7\}$
\\\hline
8 & 8 & $(-2,-7,-1)$ & $1\times \{3\}+1\times \{-4\}+1\times \{-5\}+1\times \{6\}+1\times \{-8\}+2\times \{9\}+1\times \{-10\}$ & $1\times\{1\}$
\\\hline
9 & 8 & $(-1,-4,-8)$ & $2\times \{-2\}+1\times \{-3\}+1\times \{5\}+1\times \{-6\}+1\times \{10\}+1\times \{11\}+1\times \{-13\}$ & $1\times\{8\}$
\\\hline
10 & 8 & $(-1,-8,-5)$ & $1\times \{-2\}+1\times \{-3\}+1\times \{-4\}+1\times \{-7\}+2\times \{9\}+1\times \{12\}+1\times \{-14\}$ & $1\times\{5\}$
\\\hline
11 & 8 & $(-3,-8,-1)$ & $1\times \{4\}+1\times \{-5\}+1\times \{-6\}+1\times \{7\}+1\times \{-10\}+2\times \{11\}+1\times \{-12\}$ & $1\times\{1\}$
\\\hline
12 & 8 & $(-6,-1,-2)$ & $1\times \{-3\}+1\times \{5\}+1\times \{7\}+1\times \{-9\}+1\times \{10\}+2\times \{-12\}+1\times \{14\}$ & $1\times\{2\}$
\\\hline
13 & 8 & $(-2,-5,-10)$ & $1\times \{-3\}+2\times \{-4\}+1\times \{-6\}+1\times \{7\}+1\times \{13\}+1\times \{14\}+1\times \{-17\}$ & $1\times\{10\}$
\\\hline
14 & 8 & $(-3,-10,-1)$ & $1\times \{5\}+1\times \{-6\}+1\times \{-7\}+1\times \{8\}+1\times \{-12\}+2\times \{13\}+1\times \{-14\}$ & $1\times\{1\}$
\\\hline
15 & 8 & $(-2,-5,-11)$ & $2\times \{-3\}+1\times \{-4\}+1\times \{7\}+1\times \{-8\}+1\times \{14\}+1\times \{15\}+1\times \{-18\}$ & $1\times\{11\}$
\\\hline
16 & 8 & $(-4,-1,-11)$ & $1\times \{3\}+2\times \{5\}+1\times \{6\}+1\times \{-8\}+1\times \{-14\}+1\times \{-16\}+1\times \{19\}$ &
$1\times\{11\}$
\\\hline
17 & 8 & $(-4,-11,-1)$ & $1\times \{6\}+1\times \{-7\}+1\times \{-8\}+1\times \{9\}+1\times \{-14\}+2\times \{15\}+1\times \{-16\}$ &
$1\times\{1\}$
\\\hline
18 & 8 & $(-5,-8,-7)$ & $1\times \{-3\}+1\times \{-4\}+1\times \{-6\}+1\times \{-10\}+2\times \{13\}+1\times \{17\}+1\times \{-20\}$ &
$1\times\{7\}$
\\\hline
19 & 8 & $(-4,-1,-13)$ & $2\times \{3\}+1\times \{5\}+1\times \{-8\}+1\times \{10\}+1\times \{-16\}+1\times \{-18\}+1\times \{21\}$ &
$1\times\{13\}$
\\\hline
20 & 8 & $(-8,-1,-2)$ & $1\times \{-5\}+1\times \{7\}+1\times \{9\}+1\times \{-11\}+1\times \{14\}+2\times \{-16\}+1\times \{18\}$ &
$1\times\{2\}$
\\\hline
21 & 8 & $(-2,-7,-13)$ & $1\times \{-4\}+2\times \{-5\}+1\times \{-8\}+1\times \{9\}+1\times \{17\}+1\times \{18\}+1\times \{-22\}$ & 
$1\times\{13\}$
\\\hline
22 & 8 & $(-4,-13,-1)$ & $1\times \{7\}+1\times \{-8\}+1\times \{-9\}+1\times \{10\}+1\times \{-16\}+2\times \{17\}+1\times \{-18\}$ &
$1\times\{1\}$
\\\hline
23 & 8 & $(-2,-7,-14)$ & $2\times \{-4\}+1\times \{-5\}+1\times \{9\}+1\times \{-10\}+1\times \{18\}+1\times \{19\}+1\times \{-23\}$ &
$1\times\{14\}$
\\\hline
24 & 8 & $(-5,-12,-3)$ & $1\times \{4\}+1\times \{-7\}+1\times \{-10\}+1\times \{13\}+1\times \{-14\}+2\times \{17\}+1\times \{-20\}$ &
$1\times\{3\}$
\\\hline
25 & 8 & $(-5,-14,-1)$ & $1\times \{8\}+1\times \{-9\}+1\times \{-10\}+1\times \{11\}+1\times \{-18\}+2\times \{19\}+1\times \{-20\}$ &
$1\times\{1\}$
\\\hline
26 & 8 & $(-2,-9,-15)$ & $1\times \{-4\}+2\times \{-7\}+1\times \{-8\}+1\times \{11\}+1\times \{19\}+1\times \{22\}+1\times \{-26\}$ &
$1\times\{15\}$
\\\hline
27 & 8 & $(-4,-15,-3)$ & $1\times \{5\}+1\times \{-8\}+1\times \{-11\}+1\times \{14\}+1\times \{-16\}+2\times \{19\}+1\times \{-22\}$ &
$1\times\{3\}$
\\\hline
28 & 8 & $(-2,-15,-9)$ & $1\times \{-4\}+1\times \{-5\}+1\times \{-8\}+1\times \{-13\}+2\times \{17\}+1\times \{22\}+1\times \{-26\}$ &
$1\times\{9\}$
\\\hline
29 & 8 & $(-10,-1,-2)$ & $1\times \{-7\}+1\times \{9\}+1\times \{11\}+1\times \{-13\}+1\times \{18\}+2\times \{-20\}+1\times \{22\}$ &
$1\times\{2\}$
\\\hline
30 & 8 & $(-7,-12,-9)$ & $1\times \{-4\}+1\times \{-5\}+1\times \{-10\}+1\times \{-14\}+2\times \{19\}+1\times \{23\}+1\times \{-28\}$ &
$1\times\{9\}$
\\\hline
31 & 8 & $(-11,-2,-4)$ & $1\times \{-5\}+1\times \{9\}+1\times \{13\}+1\times \{-17\}+1\times \{18\}+2\times \{-22\}+1\times \{26\}$ &
$1\times\{4\}$
\\\hline
32 & 8 & $(-12,-1,-2)$ & $1\times \{-9\}+1\times \{11\}+1\times \{13\}+1\times \{-15\}+1\times \{22\}+2\times \{-24\}+1\times \{26\}$ &
$1\times\{2\}$
\\\hline
33 & 8 & $(-8,-13,-11)$ & $1\times \{-5\}+1\times \{-6\}+1\times \{-10\}+1\times \{-16\}+2\times \{21\}+1\times \{27\}+1\times \{-32\}$ &
$1\times\{11\}$
\\\hline
34 & 8 & $(-11,-6,-12)$ & $1\times \{5\}+1\times \{7\}+1\times \{10\}+1\times \{17\}+2\times \{-22\}+1\times \{-29\}+1\times \{34\}$ &
$1\times\{12\}$
\\\hline
35 & 8 & $(-13,-2,-4)$ & $1\times \{-7\}+1\times \{11\}+1\times \{15\}+1\times \{-19\}+1\times \{22\}+2\times \{-26\}+1\times \{30\}$ &
$1\times\{4\}$
\\\hline
36 & 8 & $(-14,-1,-2)$ & $1\times \{-11\}+1\times \{13\}+1\times \{15\}+1\times \{-17\}+1\times \{26\}+2\times \{-28\}+1\times \{30\}$ & 
$1\times\{2\}$
\\\hline
37 & 8 & $(-14,-3,-6)$ & $1\times \{-5\}+1\times \{11\}+1\times \{17\}+1\times \{22\}+1\times \{-23\}+2\times \{-28\}+1\times \{34\}$ &
$1\times\{6\}$
\\\hline
38 & 8 & $(-15,-2,-4)$ & $1\times \{-9\}+1\times \{13\}+1\times \{17\}+1\times \{-21\}+1\times \{26\}+2\times \{-30\}+1\times \{34\}$ &
$1\times\{4\}$
\\\hline
\end{tabular}
}
\caption{Models with 8 or fewer Weyl fermions leading to cancellation of anomalies. Here the parameters $(a,b,c)$ appearing in the 12 fermion charges of Eq. (\ref{charge}) are taken to be integers and varied in the interval ${-15,\,15}$. A further restriction on the magnitude of the largest charge ratio is imposed, $|q_i/q_j| < 7$. Only those models with chiral charge assignments are selected. In the last column we list the scalar fields needed to generate masses for all the fermions of the respective model.}
\label{table_solutions}
\end{table}
\subsection{Scotogenic realization of minimal chiral models }\label{SEC-2.1}

\begin{figure*}[t!]
\begin{center}
\includegraphics[width=0.30\textwidth]{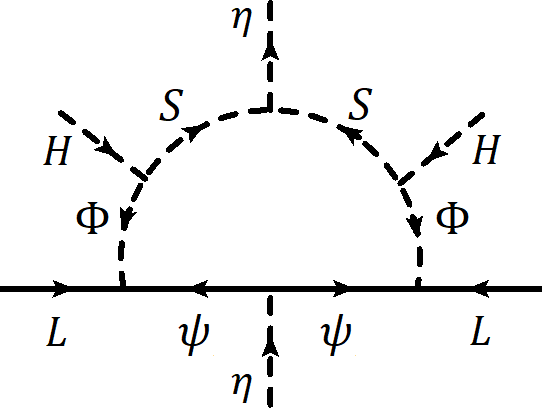}\hspace{0.19in}
\includegraphics[width=0.30\textwidth]{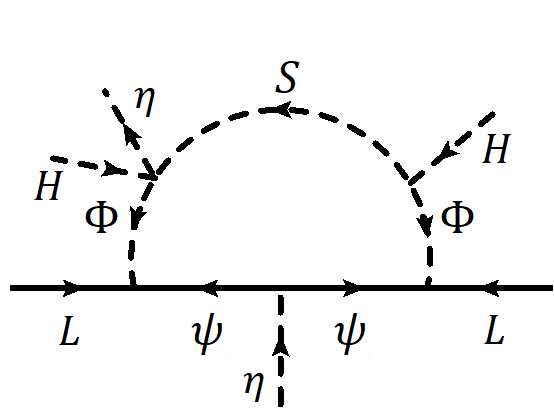} 
\hspace{0.19in}
\includegraphics[width=0.30\textwidth]{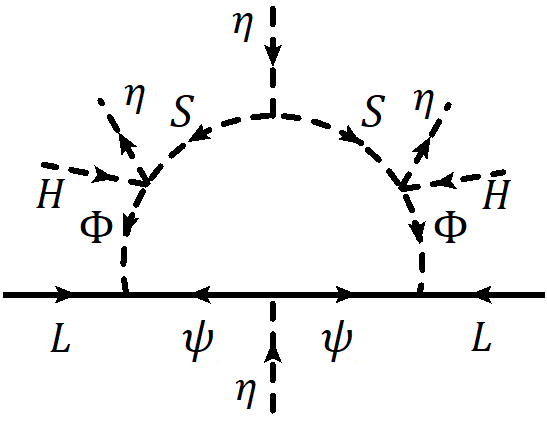} 
\end{center}
\caption{Representative Feynman diagrams for generating neutrino mass. } \label{neutmass}
\end{figure*}

In this section, we discuss the scotogenic extension of anomaly-free solutions given in Table.~\ref{table_solutions}.
To be consistent with the neutrino oscillation data,  one requires at least two heavy Majarona fermions for generating the neutrino masses. These heavy fermions can be realized within any of the anomaly-free solutions given in Table.~\ref{table_solutions}. We choose the solutions labeled Models 4 and 5 with a minimal number of scalar states to employ the scotogenic mechanism.
We find that in these two solutions, which we define as Solution-A (Model 4 in Table ~\ref{table_solutions}) and Solution-B (Model 5 in Table ~\ref{table_solutions}) with the $U(1)_D$ charges given explicitly as
\begin{align}
    &\text{Solution-A}: 2\times \{1 \} +1\times \{2 \}+ 1\times \{3 \}+ 2\times \{-4 \}+ 1\times \{-5 \}+1\times \{6 \}  \\
    &\text{Solution-B}: 1\times \{1 \}+  3\times \{2 \} + 1\times \{-3 \}+ 1\times \{-5 \}+1\times \{-6 \}+1\times \{7 \}.
\end{align}
For these two benchmark Solutions A and B, the neutrino masses are generated at one-loop level via the diagrams presented in Fig.~\ref{neutmass} and the corresponding BSM scalar states are tabulated in Table.~\ref{models}. Three scalar fields are needed to connect these loop diagrams.   

Let us comment on some particular characteristics of the rest of the model solutions presented in the Table ~\ref{table_solutions} and give our rationale for choosing Models 4 and 5 as our benchmark solutions. In the case of Model numbers 2, 3, 6, 8, 11, 14, 17, 22, and 25 in  Table ~\ref{table_solutions}, there exists no residual symmetry to stabilize the DM candidates after the spontaneous symmetry breaking of $U(1)_D$ gauge symmetry. This led us to not choose these models. Additionally, the scotogenic realization of model numbers  7, 9, 10, 12, 13, 15, 16, 18-21, 23, 24, 26-38 in Table ~\ref{table_solutions} require more than three scalar multiplets and led us to discard them on the basis of minimality of the scalar sector.

\begin{table*}[t!]
\centering
\resizebox{0.45\textwidth}{!}{
\begin{tabular}{||c|c|c||}
\hline 
Multiplets&Solution-A&Solution-B \\ \hline
$\eta$&$(1,0,2)$&$(1,0,4)$ \\
$S$&$(1,0,1)$&$(1,0,2)$ \\
$\Phi=\begin{pmatrix}
\phi^{+} \\
\phi^{0}  
\end{pmatrix}$&$(2,\frac{1}{2},-1)$&$(2,\frac{1}{2},-2)$ \\ \hline 
\end{tabular}
}
\caption{ BSM scalar sector of Solution-A and Solution-B. Here the numbers within brackets represent the quantum numbers under $SU(2)_L\times U(1)_Y \times U(1)_D$.}\label{models}
\end{table*}
For these two benchmark Solutions A and B, the stability of the DM candidate is ensured by the unbroken residual discrete symmetry surviving from the $U(1)_D$ gauge symmetry. The nature of the residual symmetry $U(1)_D \rightarrow \mathcal{Z}_N$ depends on the charge assignment of the scalar field $\eta$, which is responsible for $U(1)_D$ breaking. For Solution -A (Solution-B), the scalar field $\eta$ has a charge 2 (4) under, leading to the remnant symmetry being $\mathcal{Z}_2$ ($\mathcal{Z}_4$). 
As a case study, we present a detailed analysis of the model and framework and dark matter phenomenology for the Solution-A in the next section. We also comment on the model framework and phenomenology of Solution-B, which is very similar to that of Solution-A. 

\section{Model and framework for Solution-A}\label{SEC-03}
The benchmark case (Solution-A) that we choose to develop and analyze has the dark sector fermion charges given by \begin{align}
    &\text{Solution-A}: 2\times \{1 \} +1\times \{2 \}+ 1\times \{3 \}+ 2\times \{-4 \}+ 1\times \{-5 \}+1\times \{6 \} 
\end{align}
In addition to the SM Yukawa couplings involving the Higgs-doublet $H$, the following Yukawa terms are allowed in Solution-A:
\begin{align}
	-\mathcal{L}_Y &\supset y_{i\alpha} \overline{L}_{i}\widetilde{\Phi}\Psi_{(1)_{\alpha}}+ \frac{f_{\alpha \beta}}{2} \overline{{\Psi^{c}_{(1)}}}_{\alpha}{\Psi_{(1)}}_{\beta} \eta^* + f'_{\alpha} \overline{\Psi^c_{(2)}}{\Psi_{(-4)}}_{\alpha} \eta + f''_{\alpha} \overline{\Psi^c_{(6)}}{\Psi_{(-4)}}_{\alpha} \eta^* +  f''' \overline{\Psi^c_{(3)}}\Psi_{(-5)} \eta \nonumber \\
	&\quad +  \kappa_{\alpha} \overline{\Psi^c_{(3)}}{\Psi_{(-4)}}_{\alpha} S + \kappa' \overline{\Psi^c_{(-5)}}\Psi_{(6)} S^*  + \text{h.c.} \label{eq:Yuk}
\end{align}
where $i=e,\mu,\tau$, $\alpha,\beta=1,2$ and $\widetilde{\Phi}=i\sigma_2 \Phi^*$, with $\sigma_2$ being the second Pauli matrix. Here $\Psi_q$ represents the chiral fermion with charge $q$ under the $U(1)_D$ gauge symmetry, and $L$ is the SM lepton doublet.

After the $\eta$ field develops a vacuum expectation value (VEV), the fermion sector contains two massive Majorana fermions $\{N_1,N_2\}$ and three massive Dirac fermions $\{\xi,\zeta_1,\zeta_2\}$. The Majorana fermions acquire their masses from the Yukawa interaction term involving $f$ Yukawa matrix. We can take the matrix  $f$  to be a diagonal without loss of generality. Then the corresponding mass eigenstates can be written as $N_{\alpha}=\Psi_{(1)_{\alpha}}+\Psi^c_{(1)_{\alpha}}$, with $\alpha=1,2$. Among the three Dirac fermions, $\xi$ acquires its mass from $f'''$ Yukawa interaction term, where  $\xi=\Psi^c_{(3)}+\Psi_{(-5)}$. The remaining two Dirac fermions arise from the chiral states $\Psi_{(2)},\Psi_{(-4)_{1,2}}$, and $\Psi_{(6)}$. The corresponding mass matrix can be written as 
\begin{align}
         &  
        \begin{pmatrix}
        \Psi^c_{(2)} & \Psi^c_{(6)} 
        \end{pmatrix}
        \begin{pmatrix}
		f'_1 & f'_2   \\
		f''_1 & f''_2 
    	\end{pmatrix} 
	     \begin{pmatrix}
        \Psi_{(-4)_{1}} \\
        \Psi_{(-4)_{2}}
         \end{pmatrix}
        \langle \eta \rangle ~.
\end{align}
If for simplicity we take the mass matrix to be diagonal, then the mass eigenstates can be written as $\zeta_1=\Psi^c_{(2)}+\Psi_{(-4)_1}$ and $\zeta_2=\Psi^c_{(6)}+\Psi_{(-4)_2}$.

We note that higher dimensional operators of the type $\overline{{\Psi^{c}_{(1)}}}_{\alpha} \Psi_{(-5)}\, \eta^2/M_*$ may be added to the Yukawa Lagrangian of Eq. (\ref{eq:Yuk}), consistent with all symmetries.  If such terms are absent, the model would have a global $U(1)$ symmetry under which fermions with $U(1)_D$ charges $(3,6)$ have one unit of global charge, while those with gauge charge $(-4,-5)$ have global charge of $-1$.  Under this global $U(1)$, $\Psi_{(1)}$, $\eta$ as well as $S$, $\Phi$ and $L_i$ have zero charge. In the limit of exact global symmetry one would end up with two-component dark matter, e.g, $\Psi_{(1)}$ and the lightest of the other fermions.  While such a theory can be consistent with phenomenology, we shall assume that the higher dimensional operator noted above is present, in which case one has a single component dark matter.  This is because $\Psi_{(1)}$ will mix with $\Psi_{(-5)}$ in this case, which would allow $\Psi_{(-5)}$ to decay into $L_i + \Phi$, for example.  If the cut-off scale $M_*$ is lower than about $10^{16}$ GeV, such decays occur well before BBN, which would leave the success of big bang cosmology in tact.  In the breaking of $SU(3) \times SU(2)$ gauge symmetry down to $U(1)_D$, typically such higher dimensional operators are induced, with $M_*$ identified as this symmetry breaking scale.

The most general scalar potential of the Solution-A can be written as:
\begin{align} \begin{aligned}\label{eq:potential}
&V = -\mu^2_H H^{\dagger}H - \mu^2_\eta \eta^* \eta +\mu^2_{\Phi} \Phi^{\dagger}\Phi + \mu^2_S S^* S + (\mu_{S\Phi} H^\dagger \Phi S + \text{h.c.} ) + (\mu_{S\eta} S^2\eta^*+ \text{h.c.} ) \\ & \qquad + \sum_\varphi^{\{H,\eta,\Phi,S\}}\lambda_\varphi (\varphi^\dagger \varphi)^2 + \lambda_{H\eta} (H^{\dagger}H) \eta^{*}\eta + \lambda_{H\Phi} H^{\dagger}H \Phi^{\dagger}\Phi + \lambda'_{H\Phi} H^{\dagger}\Phi \Phi^{\dagger}H + \lambda_{\eta\Phi}  (\Phi^{\dagger}\Phi) \eta^{*}\eta
     \\ & 
\qquad + \lambda_{HS} (H^{\dagger}H) S^*S + \lambda_{\eta S} \eta^{*}\eta S^*S + \lambda_{\Phi S} (\Phi^{\dagger}\Phi) S^*S + (\lambda_{H \eta \Phi S} H^{\dagger}\Phi \eta S^* + \text{h.c.} ). 
\end{aligned}
\end{align}
All the Higgs potential parameters can be real by field rotations, which we shall assume.  The VEVs $v_H, v_D$ can be also taken to be real by gauge rotations. The electroweak (EW) symmetry is broken by the SM Higgs doublet $H$, whereas,  the $U(1)_D$ gauge symmetry is broken by the EW-singlet scalar field $\eta$. These two fields can be parameterized in the unitary gauge as
\begin{equation}
    H = \begin{pmatrix} 0 \\ 
        \frac{v_H+\sigma}{\sqrt{2}}
        \end{pmatrix}
        ,\quad \eta= \frac{v_D+\eta_r}{\sqrt{2}}.
\end{equation}
 Here $v_H\simeq246$ GeV and $v_D$ are the VEVs of the scalar field $H$ and $\eta$, respectively. In this vacuum, $U(1)_D$ gauge symmetry is broken down to discrete symmetry $\mathcal{Z}_2$, which ensures the stability of the DM candidate in this model. The mass parameters in the scalar potential and the VEVs are related by 
\begin{align}
    &\mu^2_H=\lambda_H v_H^2+\frac{\lambda_{H\eta}}{2}v_D^2,\\
    &\mu^2_{\eta}=\lambda_{\eta} v_D^2+\frac{\lambda_{H\eta}}{2}v_H^2.
\end{align}

The $\mathcal{Z}_2$ even scalar fields mix with other, the corresponding mass matrix can be written in the $\{\sigma , \eta_r\}$ basis as
\begin{align}
\mathcal{M}^2_{\sigma \eta_r}=
\begin{pmatrix}
    2\lambda_H v_H^2 & \lambda_{H\eta}v_D v_H \\
    \lambda_{H\eta}v_D v_H & 2\lambda_{\eta} v_D^2
\end{pmatrix}. \label{S0}
\end{align}
We define $h,h',$ and $\theta$ by
\begin{equation}
    \begin{pmatrix} h \\ h'\end{pmatrix}
    =\begin{pmatrix}
    \cos \theta  & -\sin \theta \\
	\sin \theta  & \cos \theta 
    \end{pmatrix}\begin{pmatrix} \sigma \\ \eta_r \end{pmatrix} 
\end{equation}
where the mixing angle $\theta$ is defined as:
\begin{equation}
    \sin 2\theta=\dfrac{2\lambda_{H\eta}v_Dv_H}{m^2_{h'}-m^2_{h}}.
\end{equation}
The field $h$ is identified as the observed Higgs boson of mass $125$ GeV. It is to be noted that this mixing angle $\theta$ is constrained from the invisible Higgs decay channel with a branching fraction of 0.145 at $95\%$ confidence level~\cite{ATLAS:2022yvh}. We shall choose our parameter set to be consistent with this limit.

In addition to these two scalar states, the model also contains one singly charged scalar  and four neutral fields, all of them transform non-trivially under the $\mathcal{Z}_2$ symmetry.
Among these four states, two of them are CP-even in nature. 
The corresponding mass-squared matrix  written in the basis $\{\text{Re}(\phi^0) , \text{Re}(S)\}$  is given by
\begin{align}
\mathcal{M}^2_{\text{Re}(\phi^0) , \text{Re}(S)}=
\begin{pmatrix}
    \mu^2_{\phi}+\frac{(\lambda_{H\Phi}+\lambda'_{H\Phi})}{2}v_H^2 + \frac{\lambda_{\eta\Phi}}{2}v_D^2 & \frac{\mu_{S\Phi}}{\sqrt{2}}v_H + \frac{\lambda_{H \eta \Phi S} }{2}v_H v_D\\
    \frac{\mu_{S\Phi}}{\sqrt{2}}v_H + \frac{\lambda_{H \eta \Phi S} }{2}v_H v_D & \mu^2_{S}+\sqrt{2}\mu_{S\eta}v_D+\frac{\lambda_{HS}}{2}v_H^2 + \frac{\lambda_{\eta S}}{2}v_D^2
\end{pmatrix}. \label{CP-even}
\end{align}
The mass eigenstates denoted as  $H_{1,2}$  are given by
\begin{equation}
    \begin{pmatrix} H_1 \\ H_2\end{pmatrix}
    =\begin{pmatrix}
    \cos \alpha  & -\sin \alpha \\
	\sin \alpha  & \cos \alpha 
    \end{pmatrix}\begin{pmatrix} \text{Re}(\phi^0) \\ \text{Re}(S) \end{pmatrix}
\end{equation}
where the mixing angle $\alpha$ is defined as 
\begin{equation}
    \sin 2\alpha=\dfrac{\sqrt{2}\mu_{S\Phi}v_H + \lambda_{H \eta \Phi S} v_H v_D}{m^2_{H_2}-m^2_{H_1}}.
\end{equation}
Here $m_{H_1}$ and $m_{H_2}$ denote the masses of $H_1$ and $H_2$, respectively.

The mass-squared matrix for the remaining  two $\mathcal{Z}_2$ odd CP-odd scalar fields can be written in the basis $\{\text{Im}(\phi^0) , \text{Im}(S)\}$  as
\begin{align}
\mathcal{M}^2_{\text{Im}(\phi^0) , \text{Im}(S)}=
\begin{pmatrix}
    \mu^2_{\phi}+\frac{(\lambda_{H\Phi}+\lambda'_{H\Phi})}{2}v_H^2 + \frac{\lambda_{\eta\Phi}}{2}v_D^2 & \frac{\mu_{S\Phi}}{\sqrt{2}}v_H +  \frac{\lambda_{H \eta \Phi S} }{2}v_H v_D\\
    \frac{\mu_{S\Phi}}{\sqrt{2}}v_H + \frac{\lambda_{H \eta \Phi S} }{2}v_H v_D& \mu^2_{S}-\sqrt{2}\mu_{S\eta}v_D+\frac{\lambda_{HS}}{2}v_H^2 + \frac{\lambda_{\eta S}}{2}v_D^2
\end{pmatrix}. \label{CP-odd}
\end{align}
The corresponding two mass eigenstates will be denoted as  $A_1$ and $A_2$ and their masses by $m_{A_1}$ and $m_{A_2}$, respectively. These states are related to the original states via the relations
\begin{equation}
    \begin{pmatrix} A_1 \\ A_2\end{pmatrix}
    =\begin{pmatrix}
    \cos \beta  & -\sin \beta \\
	\sin \beta  & \cos \beta 
    \end{pmatrix}\begin{pmatrix} \text{Im}(\phi^0) \\ \text{Im}(S) \end{pmatrix} 
\end{equation}
where the mixing angle $\beta$ is defined as
\begin{equation}
    \sin 2\beta=\dfrac{\sqrt{2}\mu_{S\Phi}v_H + \lambda_{H \eta \Phi S} v_H v_D}{m^2_{A_2}-m^2_{A_1}}.
\end{equation}

The mass of the $\mathcal{Z}_2$ odd charged scalar $\phi^{\pm}$  is given by
\begin{equation}
    m^2_{\phi^{\pm}}=\mu^2_{\Phi}+\frac{\lambda_{H\Phi}}{2}v_H^2+\frac{\lambda_{\eta\Phi}}{2}v_D^2.
\end{equation}

Once the $U(1)_D$ gauge symmetry is spontaneously broken, the associated $Z'$ gauge boson acquires a mass, which is given by 
\begin{equation}
    m_{Z'}= 2g_\chi v_\chi 
\end{equation}
The $Z'$ gauge boson interacts with the SM sector through a gauge kinetic term, which is $\mathcal{L}_{mix}\supset \frac{\epsilon}{2} F'^{\mu\nu}F_{\mu\nu}$, where $\epsilon$ is the kinetic mixing parameter, and $F_{\mu\nu}$ ($F'_{\mu\nu}$) represents the field strength tensor of the SM photon ($U(1)_D$ gauge boson) field. After the diagonalization of the kinetic terms, the interaction of $Z'$ with hidden sector and SM sector is given by
\begin{equation}
    \mathcal{L}_{mix}\supset Z'_{\mu}\left(\epsilon e \mathcal{J}^{\mu}_{\text{EM}}+g_D \mathcal{J}^{\mu}_D \right)
\end{equation}
where $\mathcal{J}^{\mu}_{\text{EM}}$ is the SM electromagnetic current and  $\mathcal{J}^{\mu}_{D}$ is the dark sector current, which is defined as 
\begin{align}
\label{eq:darkcurrent}
     \mathcal{J}^{\mu}_{D}=  \frac{1}{2}\bar{N}_{\alpha}\gamma^{\mu}\gamma_5 N_{\alpha}-\bar{\xi}\gamma^{\mu}\left(3P_L+5P_R\right)\xi-\bar{\zeta_1}\gamma^{\mu}\left(2P_L+4P_R\right)\zeta_1 -\bar{\zeta_2}\gamma^{\mu}\left(6P_L+4P_R\right)\zeta_2. 
\end{align}
Here $\alpha=1,2$, and  $P_L$ and $P_R$ denote the projection operators.
It is worth noting that in presence of the kinetic mixing, the $Z'$ gauge boson will not be stable and thus it is not a suitable DM candidate in our framework. Even though there is no kinetic mixing at the level of $SU(3) \times SU(2)$ gauge symmetry, a nonzero $\epsilon$ will be induced through loops involving the $\Phi$ scalar field which carries both $U(1)$ quantum numbers.

The neutrino mass matrix receives four contributions from Fig. \ref{neutmass} when the scalar fields involved are written in the mass eigenstates: one each from the exchange of $\Phi_\alpha= (H_1,\,H_2,\,A_1,\,A_2)$. These diagrams have overall coefficients which we define as $K_\alpha = (\cos^2\alpha,\,\sin^2\alpha,\,-\cos^2\beta,\, -\sin^2\beta)$. The masses of the Majorana fermions $(N_1,\,N_2)$, identified as $\Psi_{(1)_\alpha}$ in Eq. (\ref{eq:Yuk}), are denoted as $M_{N_{1,2}}$. We can work in a basis where the $N_{1,2}$ states are mass eigenstates. The neutrino mass matrix can then be written compactly as
\begin{eqnarray}
(m_\nu)_{ij} = \sum_{\alpha} \sum_{a=1}^2 \frac{y_{ai} y_{aj} K_\alpha}{16 \pi^2}  M_{N_a}
\left[\frac{M_\alpha^2}{M_\alpha^2 - M_{N_a}^2} {\rm log} \left(\frac{M_\alpha^2}{M_{N_a}^2} \right)  \right]
\end{eqnarray}
where the sum $\alpha$ goes over the states $\Phi_\alpha= (H_1,\,H_2,\,A_1,\,A_2)$ and where $M_\alpha$ denotes the mass of $\Phi_\alpha$. 

Since there are only two Majorana fermions circulating in the loops in Fig. \ref{neutmass}, only two neutrinos will acquire masses at one-loop.  Consequently the lightest neutrino mass is essentially massless (with higher order loop corrections inducing a very tiny mass~\cite{Babu:1988ig,Davidson:2006tg}). This testable prediction is a consequence of the scotogenic model arising from a chiral $U(1)_D$ gauge theory.

\section{Dark matter phenomenology}
\label{SEC-04}
\begin{figure}
\begin{center}
\includegraphics[width=1\textwidth]{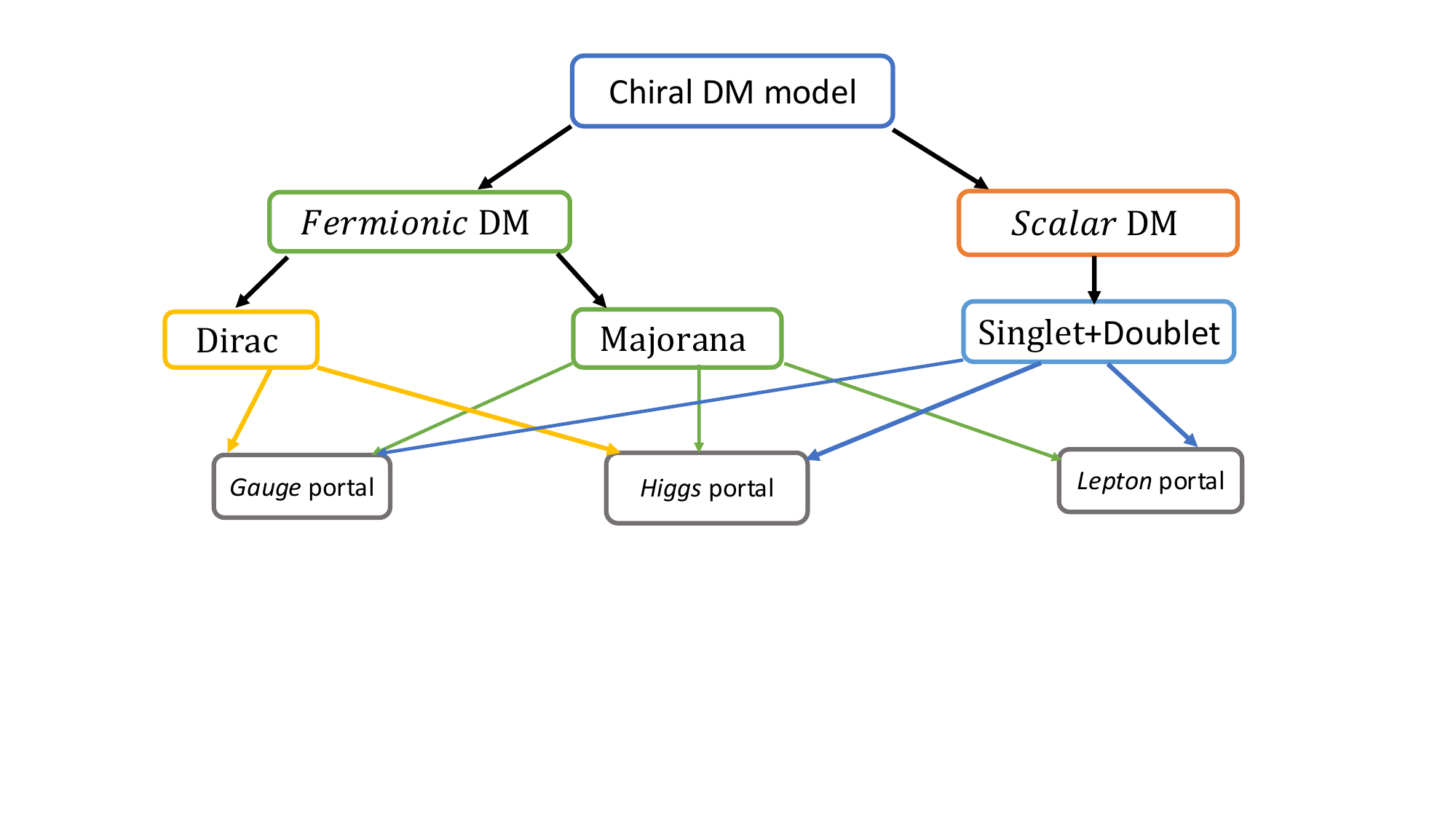}
\end{center}
\caption{Various dark matter candidates possible in the chiral DM model studied here.} \label{Flow}
\end{figure}

In this section, we discuss the DM phenomenology associated with Solution-A. As mentioned earlier, in this model, the residual $\mathcal{Z}_2$  symmetry of the $U(1)_D$ gauge symmetry stabilizes the DM candidates. The DM candidates in this framework can be either fermionic or scalar type. The various possibilities are shown in Fig. \ref{Flow}. We have shown possible portals for each case. 
We shall explore both these possibilities and the various sub-possibilities for the nature of the DM including Dirac or Majorana types for Fermionic DM or singlet+doublet type for scalar in our dark matter analysis below. Here, Dirac fermion DM arises from the additional fermions that are not involved in neutrino mass generation but take part in the anomaly cancellation. The possibility of a scalar DM arises since new scalars are required to generate neutrino masses, which could be lighter than the chiral fermions.

For the DM relic density analysis and the direct detection study, we have inserted our model in {\tt{micrOMEGAs 5.3.41}}~\cite{Belanger:2018ccd} and scanned over various parameter space. We use {\tt{LanHEP}}~\cite{Semenov:2008jy} for implementing the model. 

\subsection{Fermionic dark matter}

\begin{figure}[t!]
\centering
$$
\includegraphics[scale=0.065]{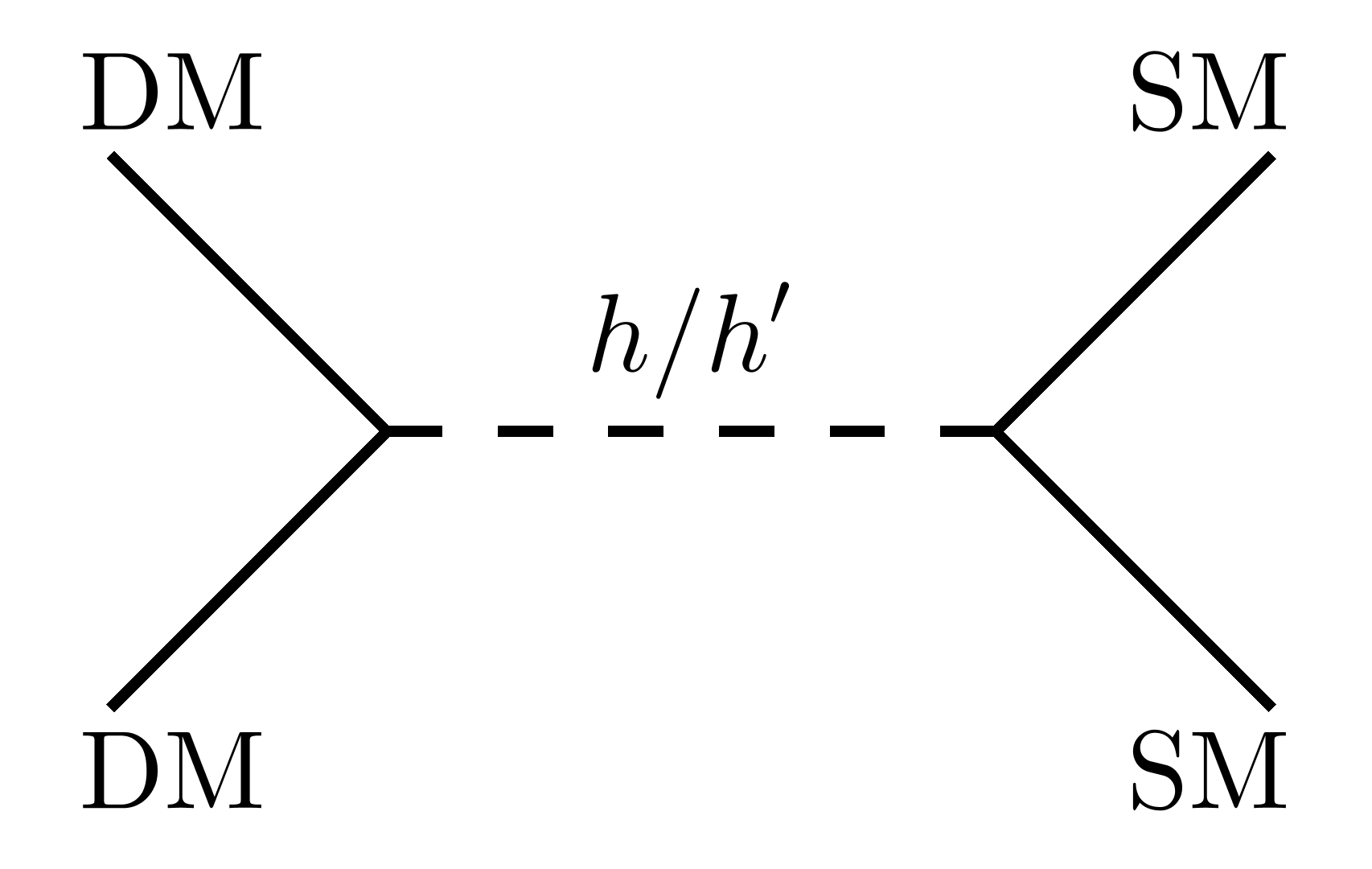}
\hspace{0.12in}
\includegraphics[scale=0.065]{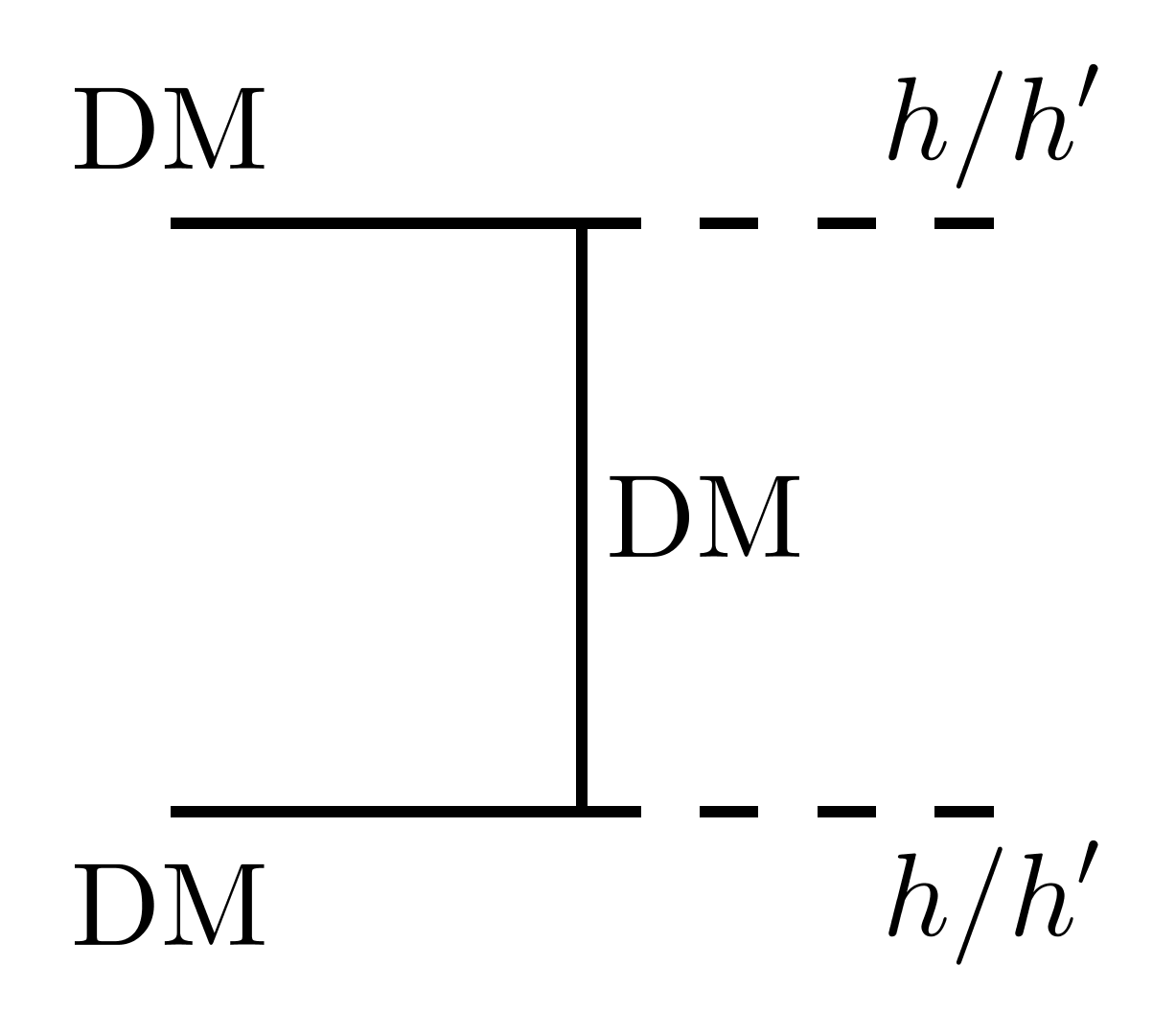}
\includegraphics[scale=0.065]{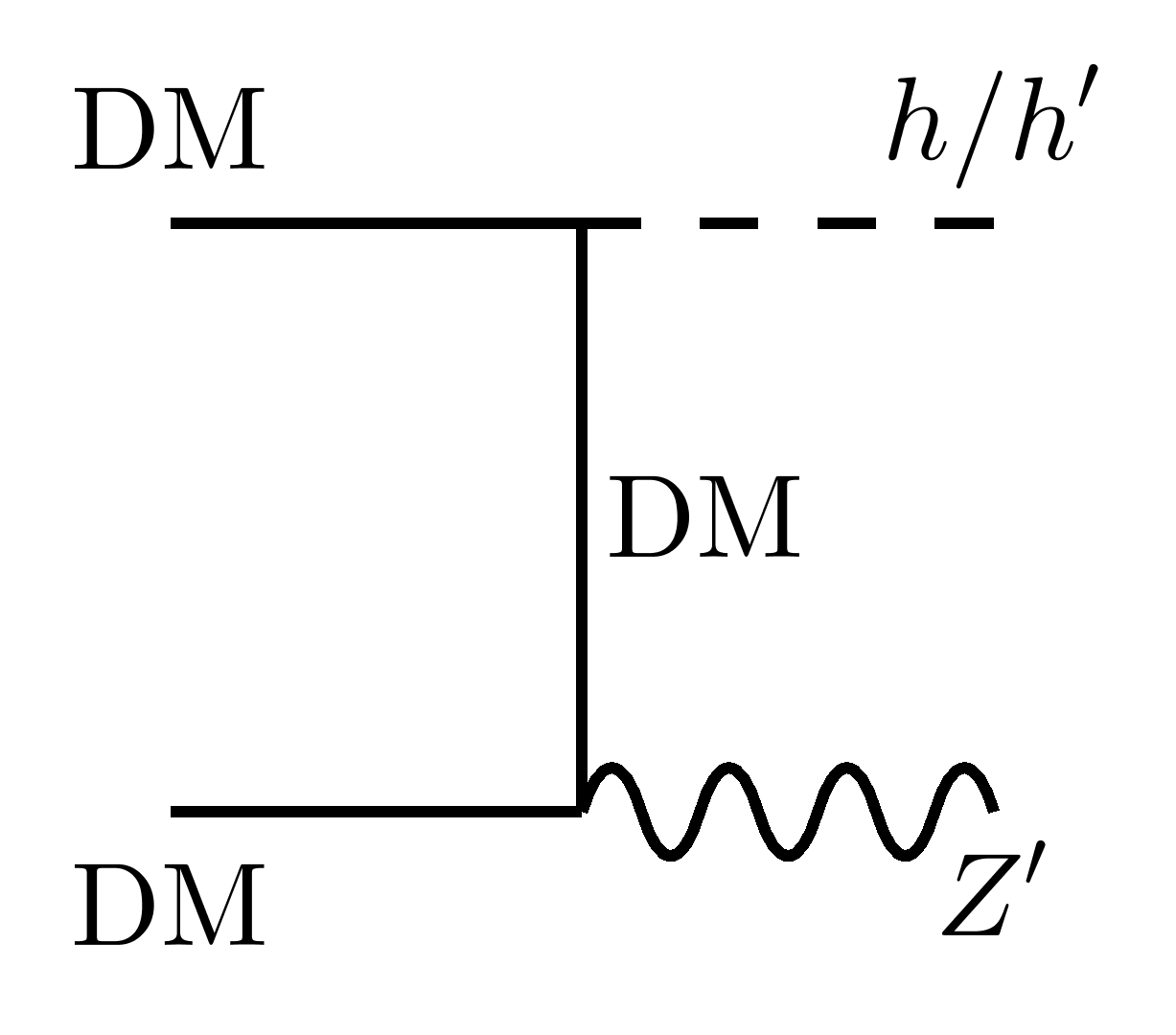}
$$
$$
\includegraphics[scale=0.065]{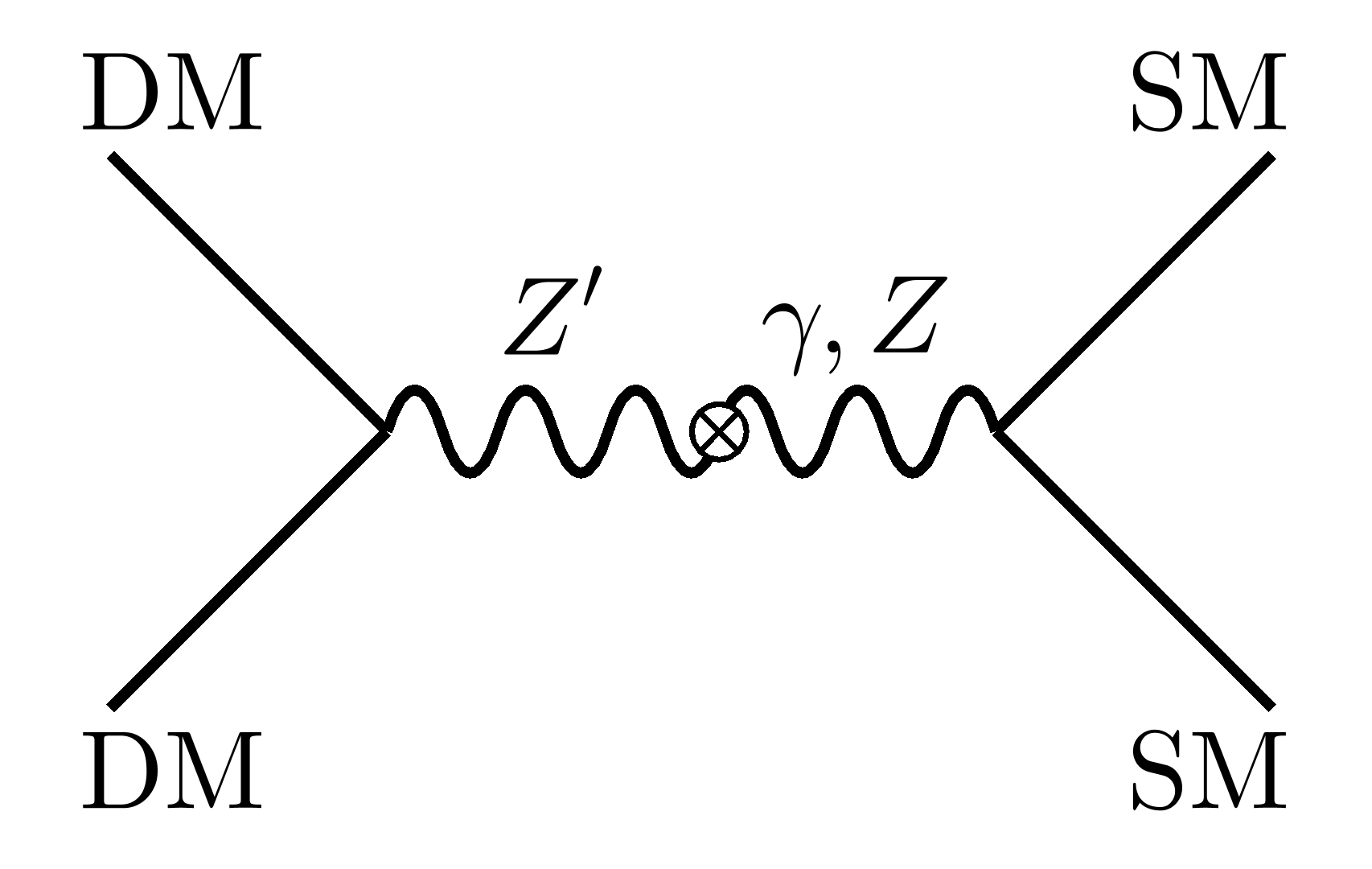}
\hspace{0.12in}
\includegraphics[scale=0.065]{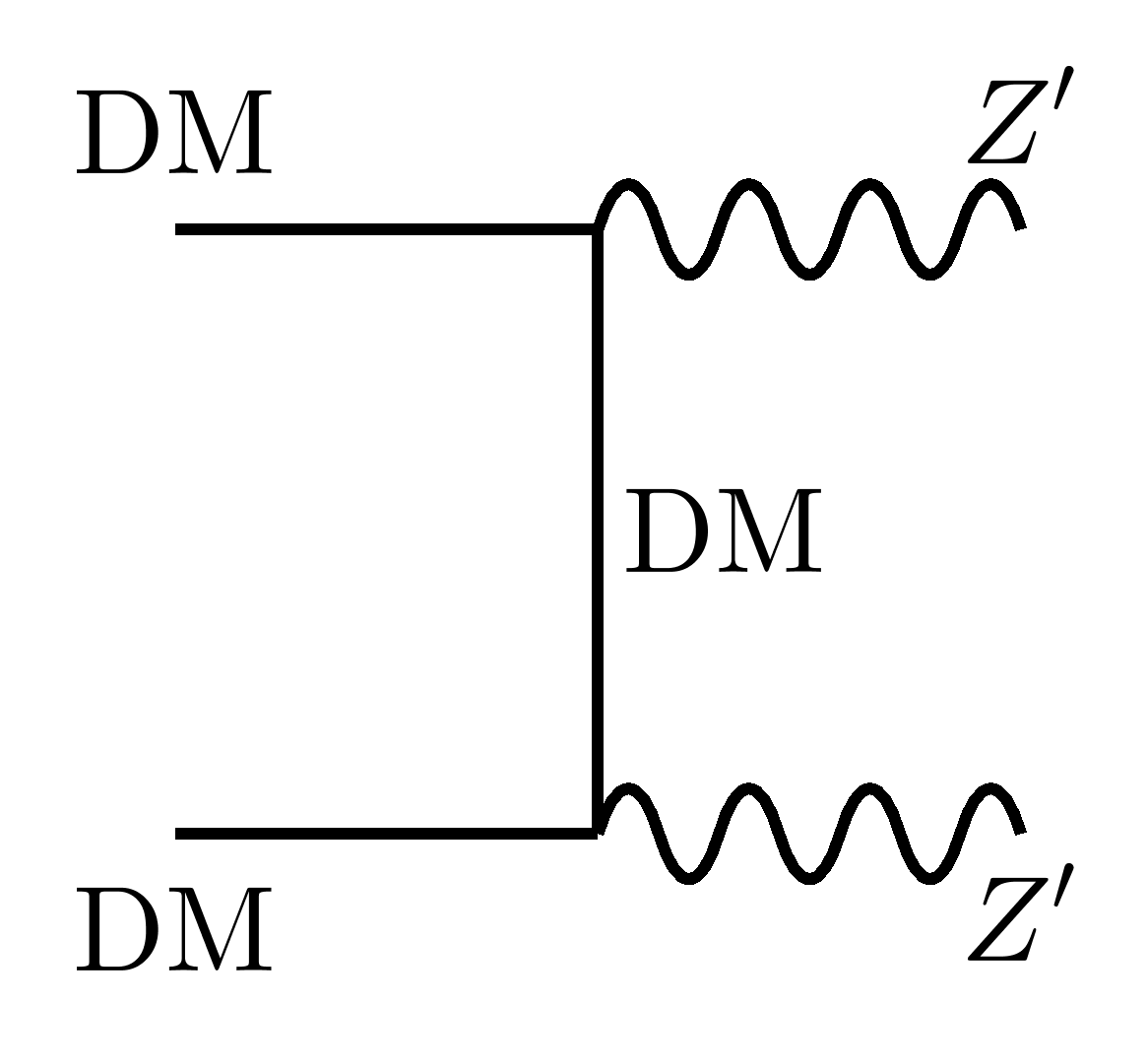}
\includegraphics[scale=0.065]{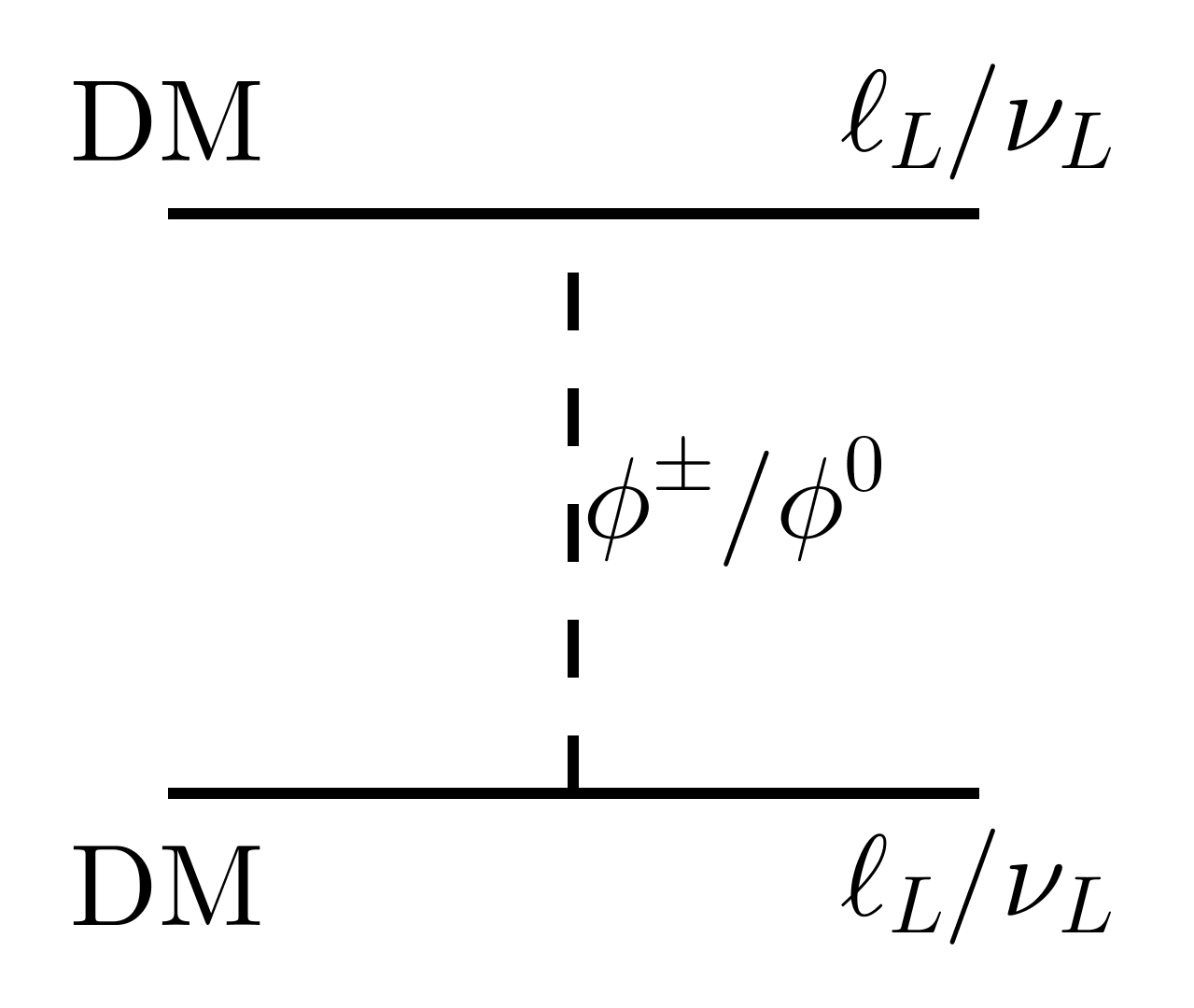}
$$
\caption{Feynman diagrams that contribute to the annihilation of Majorana DM. For Dirac dark matter, all diagrams except the last one (leptonic portal) contribute to the annihilation cross section.
}\label{FDM_ann}
\end{figure}

\begin{figure}[t!]
\centering
$$
\includegraphics[width=0.5\textwidth]{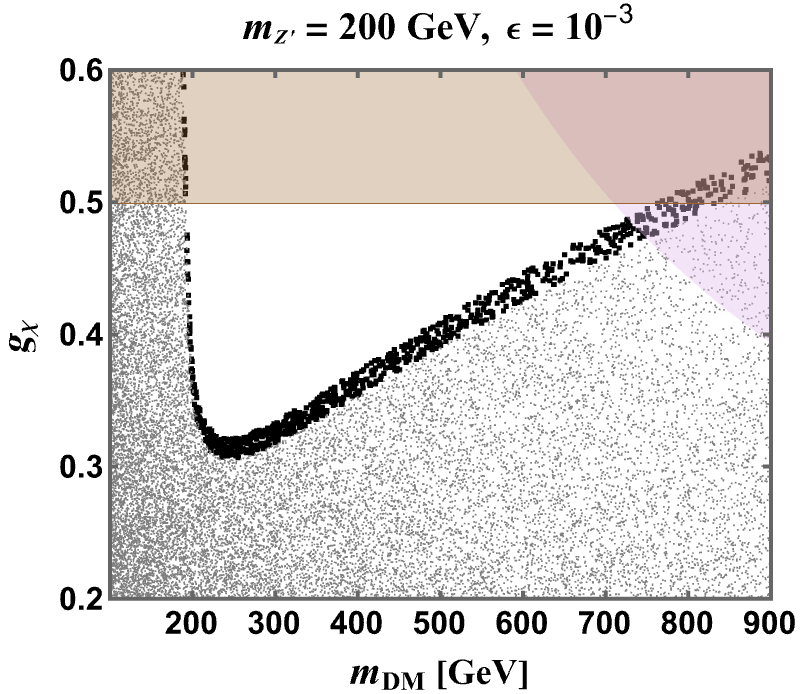}
$$
\caption{The parameter space in gauge coupling  ($g_{\chi}$) vs. DM mass ($m_{\text{DM}}$) plane consistent with the relic density constraint \cite{Planck:2018vyg} for the Majorana DM. Here the DM annihilation occurs dominantly through gauge-portal.  
The black dotted points are consistent with the relic density constraint $\Omega h^2=0.12\pm0.012$  \cite{Planck:2018vyg} for this case. The grey-colored data points are excluded from the overabundance of DM.
The brown (pink)-shaded region is excluded by the perturbativity bound on the gauge coupling (Yukawa coupling). } \label{Relic_Gauge_MajDM}
\end{figure}

\begin{figure}[t!]
\centering
$$
\includegraphics[width=0.50\textwidth]{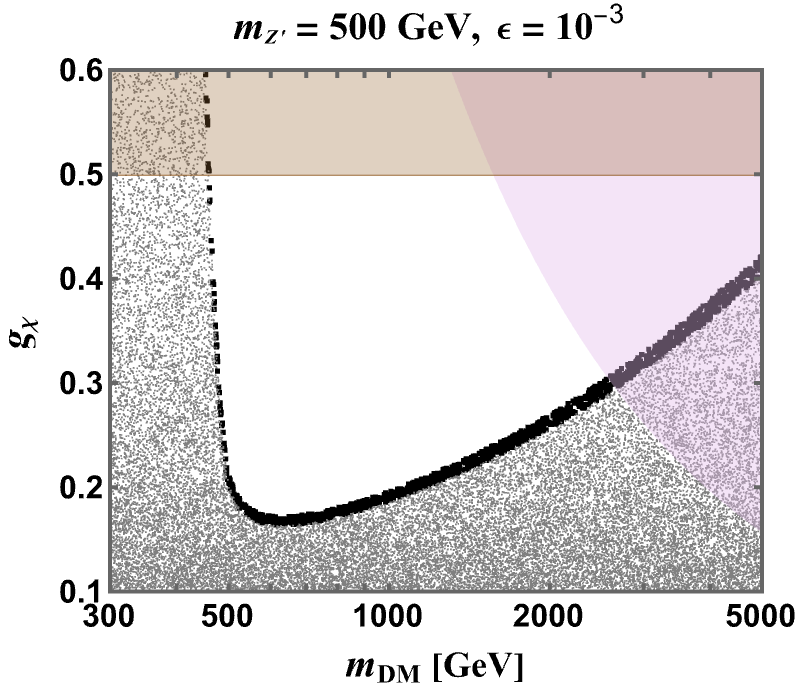}
\hspace{0.12in}
\includegraphics[width=0.5\textwidth]{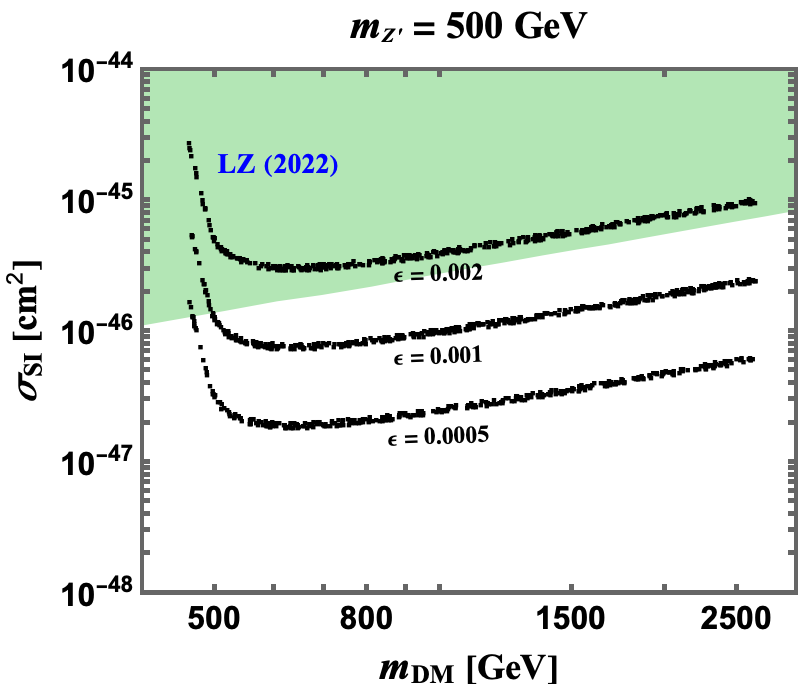}
$$
\caption{Left panel: the parameter space in gauge coupling  ($g_{\chi}$) vs. DM mass ($m_{\text{DM}}$) plane consistent with the relic density constraint \cite{Planck:2018vyg} for the Dirac DM. Here the DM annihilation occurs dominantly through gauge-portal. 
The grey-colored data points are excluded from the overabundance of DM.
The brown (pink)-shaded region is excluded by the perturbativity bound on the gauge coupling (Yukawa coupling). Right panel: DM-nucleon spin independent cross-section for different choices of $\epsilon=0.0005, 0.001$, and $0.002$. The green-shaded region denotes the excluded parameter space by the Lux-Zeplin experiment \cite{LZ:2022lsv}.} \label{Relic_Gauge_DirDM}
\end{figure}

We first focus on the fermionic DM scenario. In this case, the DM candidate can be either Majorana or Dirac fermion.  The dominant annihilation modes of the fermionic DM candidates are shown in Fig. \ref{FDM_ann}. 
First, we consider the scenario when the DM mass ($m_{\rm DM}$) $2m_{\rm DM}<m_{h',h}$. 
In this case, the DM coupling with the gauge boson $Z'$ would lead to the dominant annihilation modes for a DM pair, which are shown in the first two diagrams of the second row of Fig. \ref{FDM_ann}. For the DM mass $m_{\rm DM}> m_{Z'}$,  the dominant annihilation channel is  $\text{DM}\,+ \text{DM} \rightarrow Z'Z'$. Thus in this region of parameter space, the DM can be identified as the secluded DM \cite{Pospelov:2007mp}. 
It is important to mention that, in this scenario, for the Majorana DM case, the annihilation cross section is $p$-wave suppressed since the DM coupling with $Z'$ is axial-vector type. However, for the Dirac DM case, the DM coupling with gauge boson can be vector type, so the annihilation rate is not suppressed  and proceeds through $s$-wave.
In Fig.~\ref{Relic_Gauge_MajDM}, we show the parameter space in the $g_{\chi}$- DM mass plane and show consistency with the relic density constraint  $\Omega h^2=0.12\pm0.012$  \cite{Planck:2018vyg} as indicated by black dotted points for the Majorana DM case. The grey-colored data points correspond to $\Omega h^2>0.12+0.012$. 
Here we set the $m_{Z'}=200$ GeV and $\epsilon=10^{-3}$.
The perturbative unitarity bound the gauge coupling is given by \cite{Hally:2012pu} $g_{\chi}<0.499$, which is indicated in Fig.~\ref{Relic_Gauge_MajDM} by brown-shaded region. The corresponding bound on the Yukawa coupling $f$ is taken from \cite{Allwicher:2021rtd} $f<5.01$, which is represented as pink-shaded region in  Fig.~\ref{Relic_Gauge_MajDM}.
We see from Fig.~\ref{Relic_Gauge_MajDM} that there is a wide range of parameter space that can satisfy the relic density constraint \cite{Planck:2018vyg}.
The corresponding plot for Dirac DM $\xi$ is shown in Fig.~\ref{Relic_Gauge_DirDM} (left panel). Here we set the parameters to be $m_{Z'}=500$ GeV and $\epsilon=10^{-3}$. 
We also estimate the constraints from the
direct detection experiments \cite{XENON:2018voc, PandaX-4T:2021bab, LZ:2022lsv}, see Fig.~\ref{Relic_Gauge_DirDM} (right panel). In this scenario, the DM interacts with nucleons mainly via the exchange of $Z'$. Since the cross section of this process is proportional to the $\epsilon^2$, the gauge kinetic mixing $\epsilon$ is required to be small to satisfy the constraints from the direct detection experiments \cite{XENON:2018voc, LZ:2022lsv}.  
It is worth mentioning that, due to the absence of vectorial coupling, the direct detection constraints are much weaker for the Majorana DM.

\begin{figure}[t!]
\centering
$$\includegraphics[width=0.5\textwidth]{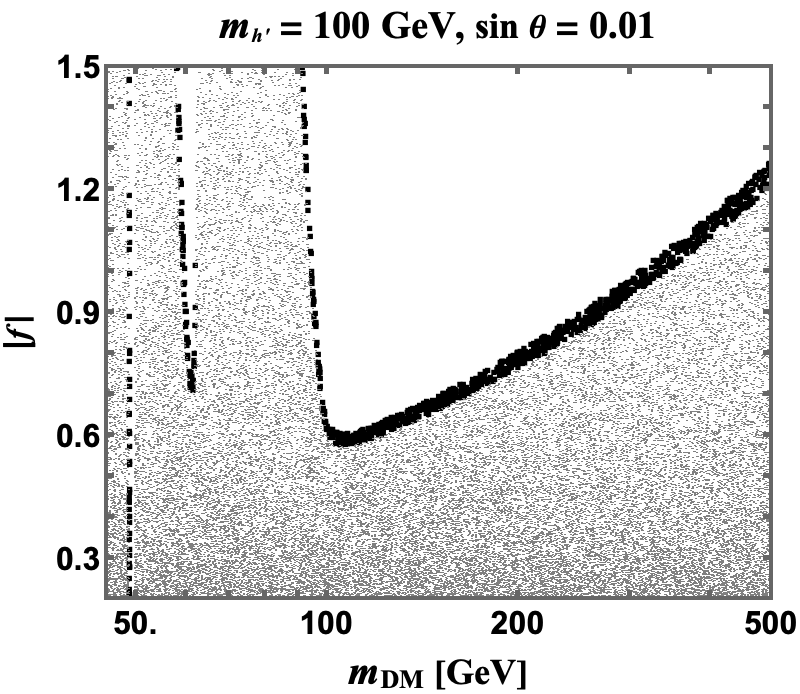}
\hspace{0.12in}
\includegraphics[width=0.52\textwidth]{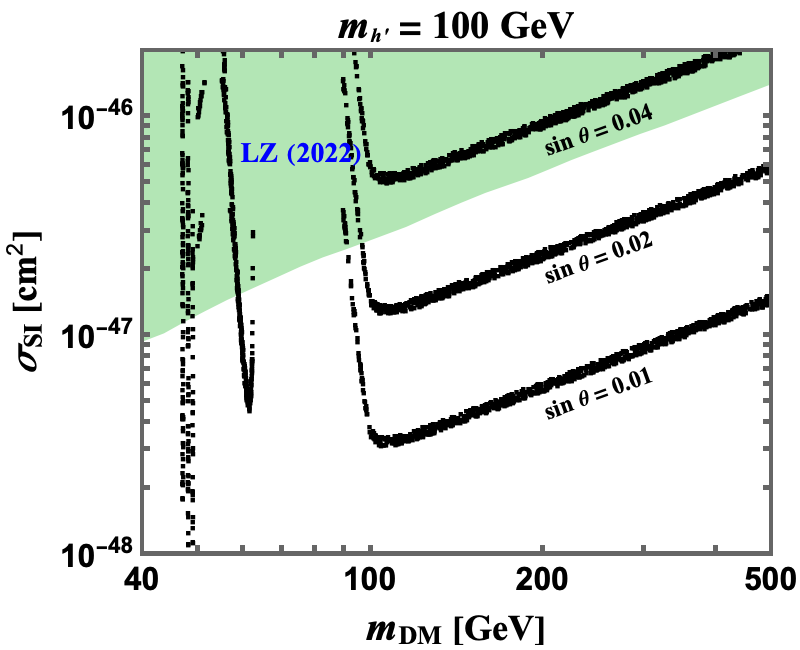}
$$
\caption{Left panel: the parameter space in Yukawa coupling  ($f$) vs. DM mass ($m_{\text{DM}}$) plane consistent with the relic density constraint \cite{Planck:2018vyg} for the Majorana DM. The grey-colored data points are excluded from the overabundance of DM. Here the DM annihilation occurs dominantly through Higgs-portal. 
Right panel: DM-nucleon spin independent cross-section for different choices of $\sin \theta=0.01, 0.02$, and $0.04$. The green-shaded region denotes the excluded parameter space by the Lux-Zeplin experiment \cite{LZ:2022lsv}.  } \label{Relic_Scalar_MajDM}
\end{figure}

Next, we investigate the case where $m_{Z'}/m_{\rm DM} \ll 1$. In this regime, the gauge coupling $g_{\chi}$ can be kept very small, making the dominant contribution to the DM annihilation cross section arise primarily through Higgs-portal channels. The corresponding annihilation modes are shown in the first two diagrams of the top row in Fig. \ref{FDM_ann}. The $s$-channel annihilation modes
can be particularly important when the mass of the DM is close to half of the scalar mediator's masses. On the other hand, when the DM is heavier than the $h$ and $h'$ fields, a DM pair can annihilate into  $h h$, $h h'$, and $h' h'$ states. However, since the DM coupling with the SM Higgs $h$ is suppressed by $\sin{\theta}$, the dominant annihilation channel is  $\text{DM}\,\text{DM} \rightarrow h'h'$. Thus, in this region of parameter space, the DM can be identified as the secluded DM \cite{Pospelov:2007mp}, and in this scenario, the limits from DM direct detection are weak. In Fig.~\ref{Relic_Scalar_MajDM} (left panel), we show the parameter space in the Yukawa coupling versus the mass of the DM plane consistent with the relic density constraint as indicated by black dotted points for the Majorana DM case.   
Here we have fixed the $m_{h'}=100$ GeV, $m_{Z'}=10^{-2} m_{\rm DM}$, $\sin{\theta}=0.01$, and $\epsilon=10^{-3}$. Notice that when the DM mass is half the $h$ or $h'$ mass, there is a resonant enhancement in the cross section, which makes otherwise disallowed region allowed. We see from Fig.~\ref{Relic_Scalar_MajDM}  that there is a wide range of parameter space where one can obtain the right amount of DM relic abundance. We have also investigated the constraints from DM direct detection experiments on the parameter space, see Fig.~\ref{Relic_Scalar_MajDM} (right panel). Here we have fixed the Yukawa couplings to obtain the measured value of DM relic abundance. In this scenario, DM interacts with nucleons via $t$-channel exchange of $h$, $h'$, and $Z'$. However, the contribution from $Z'$ is negligible due to the absence of a vectorial coupling. On the other hand, the contributions from the neutral scalar states can be large, and the cross section of these processes is proportional to  $\sin^2{2\theta}$. Hence the mixing angle should be small enough to satisfy the DM direct detection constraints. We have performed a similar analysis for the case of Dirac DM. In this case, to remain consistent with the direct detection constraints, the gauge kinetic mixing parameter $\epsilon$ must be of order $\mathcal{O}(10^{-7})$, due to the vectorial coupling between Dirac dark matter and $Z'$, as shown in Eq. (\ref{eq:darkcurrent}). However, achieving such small values of $\epsilon$ requires significant fine-tuning.

\begin{figure}[t!]
\centering
$$\includegraphics[width=0.5\textwidth]{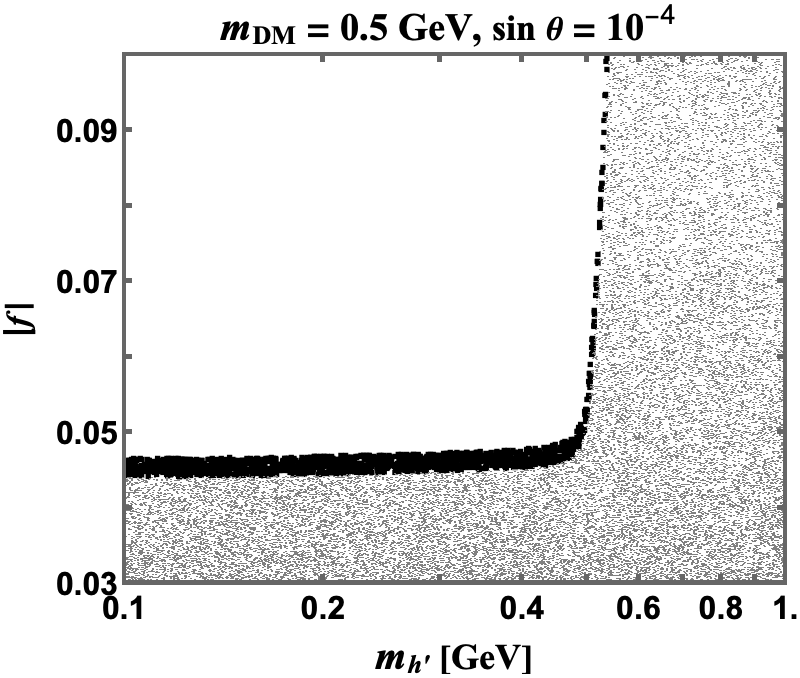}
\hspace{0.12in}
\includegraphics[width=0.5\textwidth]{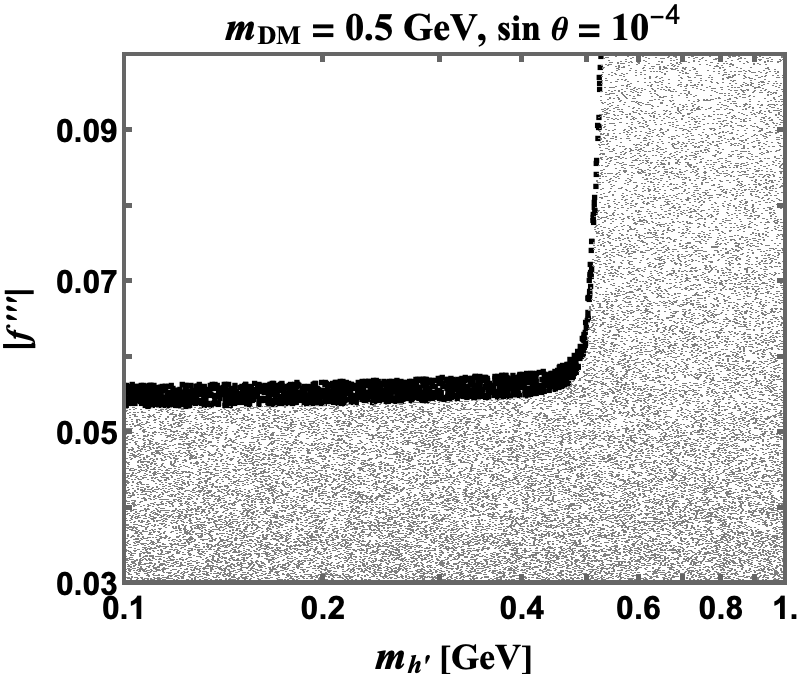}
$$
\caption{The parameter space in Yukawa coupling vs. DM mass plane consistent with the  relic density constraint \cite{Planck:2018vyg} for both Majorana (left) and Dirac DM (right).} \label{LightDM}
\end{figure}

In addition to the heavy DM case, this model could also accommodate a light DM scenario.   To illustrate this, in Fig.~\ref{LightDM}, we show the parameter space for $m_{\rm DM}=500$ MeV for both Majorana and Dirac DM. Here we have chosen $m_{h'}<m_{\rm DM}$ and 
$\sin \theta = 10^{-4}$. In this case, the DM mainly annihilates into $h'h'$, hence avoiding the stringent constraints imposed by the CMB observables \cite{Leane:2018kjk}. 
Constraints from the meson decays can be relevant for such a light scalar. However, for a mixing angle $\sin \theta$ is of the $\mathcal{O}(10^{-4})$, these constraints can be easily satisfied \cite{Dev:2024twk}.

\subsection{Scalar dark matter}
\begin{figure}[t!]
\centering
$$
\includegraphics[scale=0.3]{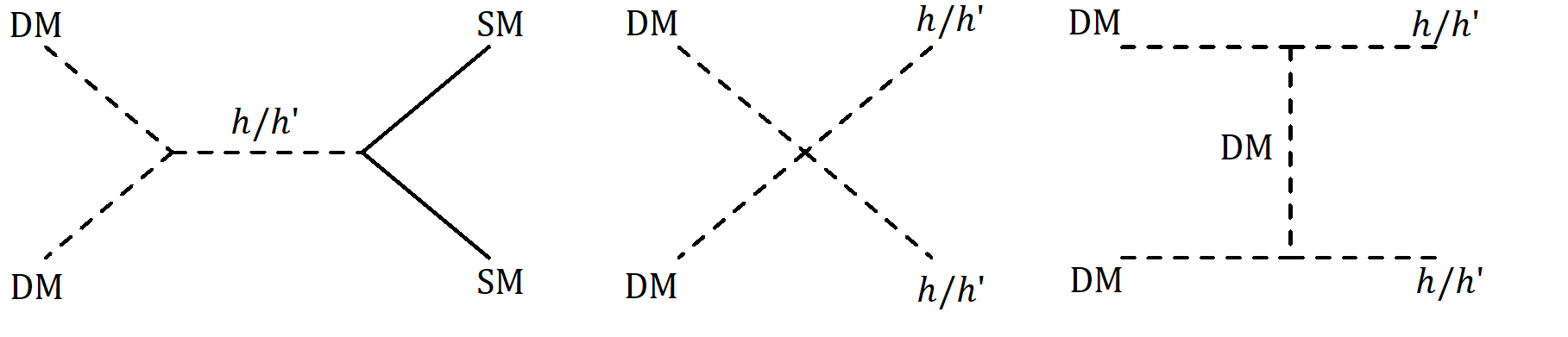}
$$
$$
\includegraphics[scale=0.3]{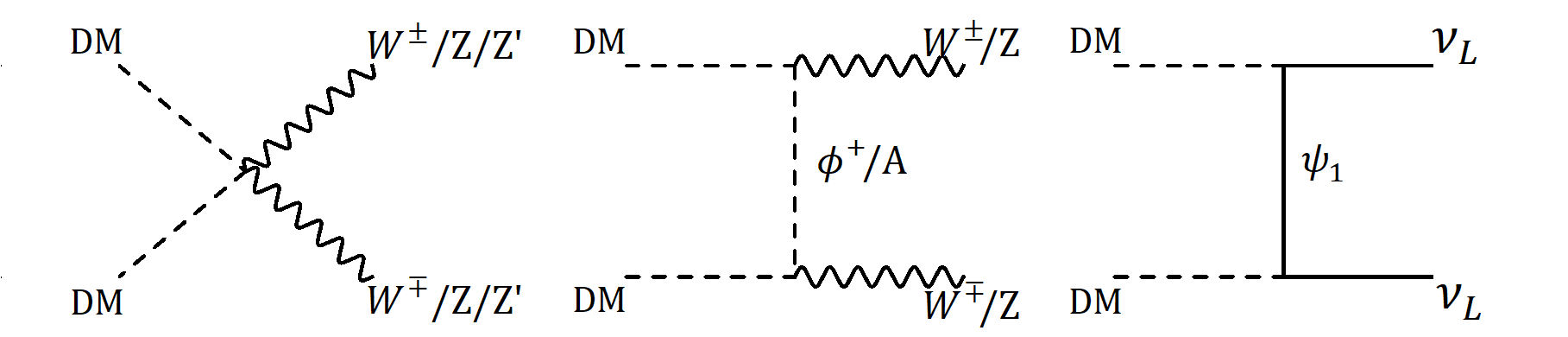}
$$
\caption{Feynman diagrams that contribute to the annihilation of the scalar DM.
}\label{SDM_ann}
\end{figure}

In this subsection, we discuss the DM phenomenology associated with scalar DM.  In order to generate neutrino mass, one needs to have a nonzero mixing between scalar doublet $\Phi$ and the singlet scalar $S$. Hence, the scalar DM in our setup will be an admixture of neutral components of  $\Phi$ \cite{PhysRevD.18.2574, Barbieri:2006dq} and  $S$ \cite{Silveira:1985rk,McDonald:1993ex}. The lightest neutral state among the $\{H_1,H_2,A_1,A_2\}$ qualifies to be a DM candidate. For the DM analysis,  we take $H_1$ to be the DM candidate. A pair of scalar DM $H_1$ can annihilate through modes, which are shown in Fig.~\ref{SDM_ann}. In addition to these channels, the co-annihilation of DM with other $Z_2$-odd particles, as well as the annihilation of the co-annihilation partners,  could play a crucial role in DM phenomenology, particularly when the mass difference between DM and the co-annihilation partners is small \cite{Griest:1990kh}. With the help of {\tt{micrOMEGAs}}~\cite{Belanger:2018ccd}, we incorporate all these relevant annihilation channels in our study.

\begin{figure}[t!]
\centering
\centering
$$\includegraphics[width=0.5\textwidth]{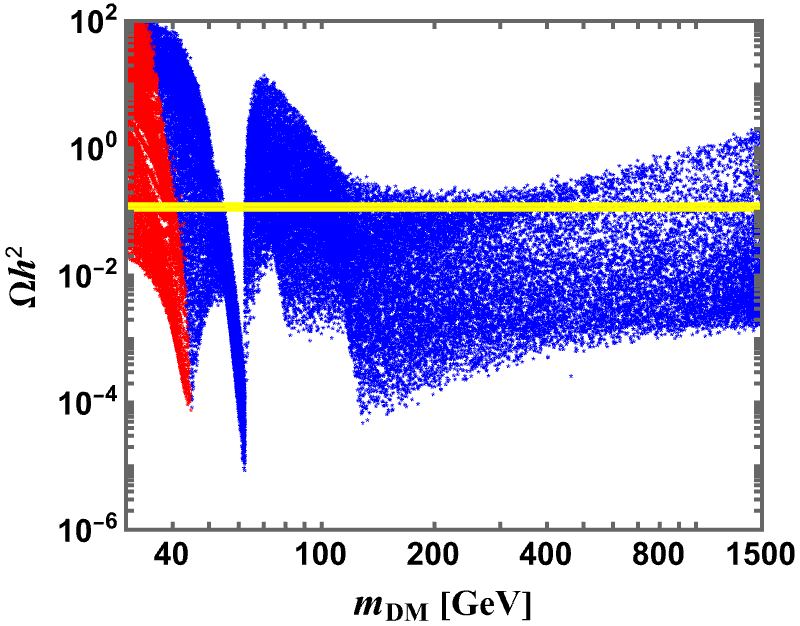}
\hspace{0.12in}
\includegraphics[width=0.5\textwidth]{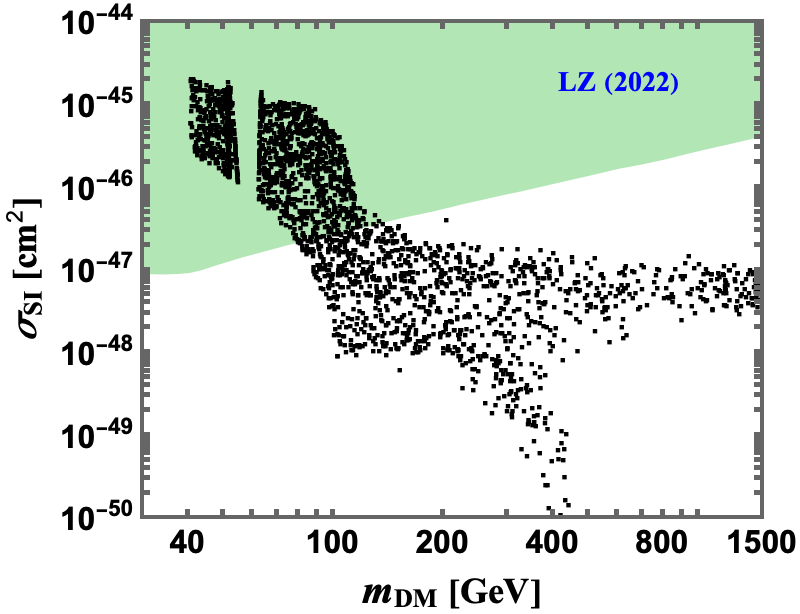}
$$
\caption{Left panel: Relic density of DM as a function of $m_{\rm DM}$. The red (blue)-colored data points are excluded (allowed) by the constraints from the $Z$-decay width measurements~\cite{Electroweak:2003ram,ParticleDataGroup:2024cfk}. The yellow-shaded region indicates the Planck relic density constraints \cite{Planck:2018vyg}. Right panel: DM-nucleon spin independent cross section. The green-shaded region denotes the excluded parameter space by the Lux-Zeplin experiment \cite{LZ:2022lsv}.
Here, we set $\sin \alpha=0.99$.} \label{ScalarDM}
\end{figure}

In Fig.~\ref{ScalarDM} (left panel), we show the parameter space in DM relic density versus $m_{\rm DM}$ plane for $\sin \alpha =0.99$. For this, we randomly scan over the ranges $m_{H_1}<m_{H_2},m_{A_1},m_{A_2}\in (10 \, \text{GeV} - 2000 \,\text{GeV})$, $m_{\phi^{\pm}}\in (110 \, \text{GeV} - 2000 \,\text{GeV})$, and $\sin\beta \in (10^{-3}, 0.99)$. We fix the quartic couplings $\lambda_{H\Phi},\lambda_{HS},\lambda_{\eta \Phi},\lambda_{\eta S}, \lambda_{H\eta \Phi S}\simeq 0$ and choose the mixing angle $\sin \theta \simeq 0$. From this set of data points, we select the parameters that satisfy the perturbative unitarity constraints on the quartic couplings ($\lambda<4\pi$ \cite{Casalbuoni:1986hy})  and also satisfy the constraints imposed by the electroweak precision data \cite{ParticleDataGroup:2024cfk}.  The red (blue)-colored data points are excluded (allowed) by the constraints from the $Z$-decay width measurements~\cite{Electroweak:2003ram,ParticleDataGroup:2024cfk}. In Fig.~\ref{ScalarDM} (right panel), we show the limits imposed by the Lux-Zeplin experiment \cite{LZ:2022lsv}.

\subsection{Brief remarks on model based on Solution-B}

While our phenomenological analysis of Sec. \ref{SEC-04} focused on the model based on Solution-A, this analysis can be extended to Solution-B in a straightforward way.  We note that the Higgs potential of Solution-B is identical to that of Solution-A, since the scalar fields have the same gauge quantum number, except for the fact that the $U(1)_D$ charges are twice as large as in Solution-A, see Table \ref{models}. The Higgs potential analysis is thus the same as in Solution-A. The Yukawa couplings for Solution-B are given by the Lagrangian  
\begin{align}
	-\mathcal{L}_Y &\supset y_{i\alpha} \overline{L}_{i}\widetilde{\Phi}\Psi_{(2)_{\alpha}}+ \frac{f_{\alpha \beta}}{2} \overline{{\Psi^{c}_{(2)}}}_{\alpha}{\Psi_{(2)}}_{\beta} \eta^* + f' \overline{\Psi^c_{(-3)}}{\Psi_{(7)}} \eta^* + f''_{\alpha} \overline{\Psi^c_{(2)_\alpha}}{\Psi_{(-6)}} \eta  + \overline{\Psi^c_{(1)}}{\Psi_{(-5)}} \eta \nonumber \\
	&\quad +    \kappa \overline{\Psi^c_{(1)}}{\Psi_{(1)}} S^* + \kappa' \overline{\Psi^c_{(-1)}}\Psi_{(-3)} S  + \text{h.c.} \label{eq:Yuk2}
\end{align}

This structure is very similar to Eq. (\ref{eq:Yuk}), and will generate a $4 \times 4$ Majorana mass matrix for $(\Psi_{(2)_\alpha},\, \Psi_{(-6)}$ sector and  two  Dirac masses connecting $(\Psi_{(-3)},\, \Psi_{(7)})$ and $(\Psi_{(1)},\, \Psi_{(-5)})$ . As in Solution-A, these sectors will develop cross couplings via higher dimensional operators which can result in single-component dark matter. In this case, since there are three copies of Majorana fermions $\Psi_2$ that take part in the scotogenic mechanism for neutrino masses, all light neutrinos will become massive, unlike Solution-A, where one light neutrino remains massless.

\section{Conclusion}
\label{SEC-05}

In this paper we have proposed a class of chiral  models for dark matter based on a $U(1)_D$ gauge symmetry acting on a dark sector.  Owing to its chiral nature the $U(1)$ protects the mass of the dark matter which arises after spontaneous symmetry breaking.  An unbroken $\mathcal{Z}_N$ discrete symmetry with its origin in the gauge $U(1)_D$ ensures the stability of the dark matter in this scenario, making this class of models appealing.  

Our approach to generating the chiral gauge models is to start with a simple and chiral non-Abelain gauge symmetry and reduce it down to $U(1)_D$.  For the non-Abelian gauge symmetry we chose $SU(3) \times SU(2)$ with one family of fermions in the dark sector which resembles the SM fermions.  Chiral models with a smaller number of Weyl fermions are identified by removing certain fermion pairs which have vectorial charges under the surviving $U(1)_D$. With simplifying restrictions on the charges of the fermions, we have identified 38 models which are listed in Table \ref{table_solutions}.  In all but 3 of these models a single scalar field is sufficient to generate masses for all dark sector fermions. Furthermore, we have used these dark sector fermions to generate neutrino masses via the scotogenic mechanism by enhancing the scalar sector of these models.

We have carried out a detailed phenomenological analysis of dark matter in one of the simplest models in this class.  Once neutrino masses are induced within this framework, the model allows for three possible dark matter candidates: a Majorana fermion, a Dirac fermion, or a singlet+doublet scalar.  We have shown the consistency of each of these scenarios with dark matter relic abundance as well as direct detection limits and found that for some of the parameters of the model the dark matter candidate should be accessible to the direct detection experiments.  

\section*{Acknowledgments}
The work of KSB was supported in part by the United States Department of Energy Grant No. DE-SC0016013. The work of SC is supported by the College of the Holy Cross Robert L. Ardizzone (’63) Fund for Tenure Track Faculty Excellence. The authors acknowledge the Center for Theoretical Underground Physics and Related Areas (CETUP*) and the Institute for Underground
Science at Sanford Underground Research Facility (SURF) for providing a conducive environment during the summer workshops in 2023 and 2024 where part of this work was done. We like to thank Bhupal Dev for helpful discussions.

\bibliographystyle{utphys}
\bibliography{reference}

\end{document}